\documentclass[journal=jacsat,manuscript=article]{achemso}

\usepackage[version=3]{mhchem} 

\usepackage[version=3]{mhchem} 
\usepackage{acronym} 
\usepackage{textcomp, gensymb}
\usepackage{amssymb}
\usepackage{diagbox}
\usepackage{multirow}
\usepackage{graphicx}
\usepackage{amsmath}
\usepackage{xcolor}
\usepackage{array,booktabs,tabularx}    
\usepackage{hyperref} 
\usepackage{float}

\acrodef{MD}[MD]{molecular dynamics}
\acrodef{DFT}[DFT]{density functional theory}
\acrodef{LIB}[LIB]{lithium-ion battery}
\newacroplural{LIB}[LIBs]{lithium-ion batteries}
\acrodef{VASP}[VASP]{Vienna \textit{ab-initio} simulation package}
\acrodef{PBE}[PBE]{Perdew-Burke-Ernzerhof}

\acrodef{SIB}[SIB]{sodium-ion battery}
\acrodef{PBA}[PBA]{Prussian blue analogue}

\acrodef{RAZIB}[RAZIB]{rechargeable aqueous zinc-ion battery}
\newacroplural{RAZIB}[RAZIBs]{Rechargeable aqueous zinc-ion batteries}
\acrodef{GGA}[GGA]{generalized gradient approximation}
\acrodef{OER}[OER]{oxygen evolution reaction}
\acrodef{HER}[HER]{hydrogen evolution reaction}
\acrodef{SHE}[SHE]{standard hydrogen electrode}
\acrodef{elect_pot}[\textit{E}]{electrode potential}

\acrodef{VRE}[VRE]{variable renewable energy}
\acrodef{GSES}[GSES]{grid-scale energy storage}
\acrodef{NASICON}[NASICON]{sodium super ionic conductor}

\acrodef{NEB}[NEB]{nudged elastic band}
\acrodef{ASE}[ASE]{atomic simulation environment}
\acrodef{API}[API]{application programming interface}
\acrodef{DOS}[DOS]{density of states}
\acrodef{PDOS}[pDOS]{projected density of states}
\acrodef{COHP}[COHP]{crystal orbital Hamilton population}
\acrodef{pCOHP}[pCOHP]{projected crystal orbital Hamilton population}
\acrodef{-IpCOHP}[-IpCOHP]{negative integrated projected crystal orbital Hamilton population}
\acrodef{Ef}[E$_{\text{F}}$]{Fermi energy}

\acrodef{percdist}[\textit{d}$_\text{perc}$]{percolation path distance}
\acrodef{percdistmax}[\textit{d}$_\text{perc}^\text{max}$]{maximum percolation path distance}
\acrodef{percdisp}[$\delta_\text{perc}$]{maximum percolation path deviation}

\acrodef{dGpbx}[$\Delta$\textit{G}$_\text{pbx}$]{electrochemical decomposition energy}
\acrodef{avgdGpbx}[$\langle\Delta$\textit{G}$_\text{pbx}\rangle$]{average electrochemical decomposition energy}

\acrodef{EZn}[\textit{E}$_\text{Zn}$]{Zn$^{2+}$ (de)intercalation potential}
\acrodef{EZninitial}[\textit{E}$^{i}_\text{Zn}$]{initial Zn$^{2+}$ (de)intercalation potential}
\acrodef{EZnavg}[$\langle$\textit{E}$_\text{Zn}\rangle$]{average Zn$^{2+}$ (de)intercalation potential}

\acrodef{HZn}[$\Delta$\textit{H}$_\text{Zn}$]{Zn$^{2+}$ (de)intercalation energy}
\acrodef{mchg}[$\text{M}_\text{chg}$]{charged battery state}
\acrodef{mdchg}[$\text{M}_\text{dchg}$]{discharged battery state}
\acrodef{Etot}[\textit{E}$_\text{tot}$]{total energy}

\acrodef{Qw}[\textit{Q}$_\text{w}$]{theoretical gravimetric capacity}
\acrodef{Ww}[\textit{W}$_\text{w}$]{theoretical energy density}

\acrodef{XRD}[XRD]{X-ray diffraction}
\acrodef{PVDF}[PVDF]{polyvinylidene fluoride}
\acrodef{NMP}[NMP]{N-Methyl-2-pyrrolidone }
\acrodef{CV}[CV]{cyclic voltammetry}
\acrodef{GCD}[GCD]{galvanostatic charge-discharge}

\acrodef{overpotential}[$\eta$]{overpotential}
\acrodef{overpotential_OER}[$\eta_{\text{OER}}$]{overpotential associated with the \ac{OER}}

\acrodef{Eb}[$E_{\text{b}}$]{percolation energy barrier}

\acrodef{SI}[SI]{Supporting Information}

\newcommand{\titlename}{Computational discovery of cathode materials for rechargeable aqueous zinc-ion batteries}

\author{Caio Miranda Miliante}
\email{miliantc@mcmaster.ca}
\affiliation{Department of Materials Science and Engineering, McMaster University, 1280 Main Street West, Hamilton, Ontario L8S 4L8, Canada}
\author{Yuzhen Deng}
\affiliation{Department of Chemical Engineering, McMaster University, 1280 Main Street West, Hamilton, Ontario L8S 4L8, Canada}

\author{Brian D. Adams}
\affiliation{Salient Energy Inc., Dartmouth, Nova Scotia B3B 1C4, Canada}
\author{Drew Higgins}
\affiliation{Department of Chemical Engineering, McMaster University, 1280 Main Street West, Hamilton, Ontario L8S 4L8, Canada}
\author{Oleg Rubel}
\affiliation{Department of Materials Science and Engineering, McMaster University, 1280 Main Street West, Hamilton, Ontario L8S 4L8, Canada}

\title{\titlename}

\begin{document}

\begin{tocentry}




\includegraphics[]{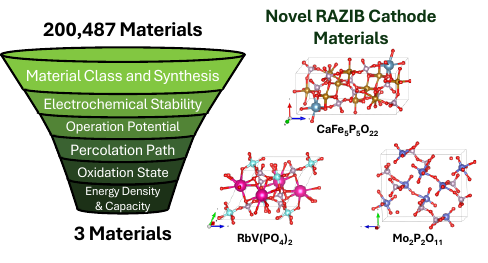}

\end{tocentry}

\begin{abstract}
    \Acp{RAZIB} attract considerable scientific and commercial interest for deployment in grid-scale energy storage due to higher safety and lower manufacturing cost when compared to lithium-ion batteries. However, currently studied cathode materials suffer from severe capacity fade when cycling at rates appropriate for grid-scale applications ($<$~C/2), which hampers the commercialization of \acp{RAZIB}. To address the present limitation on cathode material availability, more than 2000 previously synthesized oxides, chalcogenides, Prussian blue analogues, and polyanion materials were computationally screened for the discovery of highly stable \ac{RAZIB} cathode materials. The structural, electrochemical, and chemical properties of the materials were respectively evaluated through an investigation of the available \ce{Zn^{2+}} percolation paths in the crystal structure, the stability of the material in aqueous media under \ac{RAZIB} operation conditions, and the attained transition metal oxidation state during cycling. The transition metal oxidation state and intercalating ion coordination environment were determined to govern the magnitude of the calculated \ce{Zn^{2+}} intercalation potential, with this finding being useful to guide the development of batteries with high operating voltages. Additionally, the bonding character of \ce{Fe}-\ce{O} bonds was contrasted between olivine-\ce{FePO4} and $\alpha$-\ce{FePO4}. The latter material was successfully synthesized but was found to be  electrochemically inactive as a cathode for \acp{RAZIB}. Finally, three previously synthesized materials, but unexplored as cathodes for \acp{RAZIB} (\ce{CaFe5P5O22}, \ce{RbV(PO4)2}, and \ce{Mo2P2O11}), were identified for the first time to have high \ce{Zn^{2+}} (de)intercalation potential, electrochemical stability, theoretical gravimetric capacity, and energy density, and are proposed here for further experimental testing. Overall, the results presented in this study present a guide for the experimental development of stable next-generation cathode materials for \acp{RAZIB}, and help expedite the establishment of \ac{RAZIB} as a commercially viable technology for grid-scale energy storage.
\end{abstract}

\section{Introduction}

Rechargeable batteries have been frequently deployed as grid-scale energy storage infrastructure in support of the energy transition to renewable sources, as batteries have a modular design, smaller footprint, and lower cost\cite{Zhu_CR_122_2022, Gallo_RSER_65_2016}. \Acfp{RAZIB}, for example, are a promising battery technology for grid-scale applications, being considered significantly safer than potentially flammable \acp{LIB} due to the use of an aqueous electrolyte\cite{Gourley_Joule_7_2023, Gupta_AEM_15_2025, Fan_PMS_149_2025, Li_AFM_34_2024}. Also, \acp{RAZIB} have a lower associated manufacturing cost and utilizes a more abundant and non-toxic metal species in \ce{Zn}, which would allow for easier scaling from lab to commercial deployment\cite{Gourley_Joule_7_2023, Li_AFM_34_2024, Fan_PMS_149_2025}. However, currently researched cathode materials for \ac{RAZIB} suffer from severe capacity fade, especially when cycling at practical rates for grid-scale operation (\textit{i.e.}, C-rates lower than C/2), which directly hampers the \ac{RAZIB} commercialization efforts\cite{Gourley_Joule_7_2023, Miliante_JPCC_128_2024, Li_AFM_34_2024, Gupta_AEM_15_2025}.

Both organic and inorganic materials have been experimentally investigated as cathode materials for \acp{RAZIB}. Organic materials draw considerable interest in the literature as their structural diversity allow researchers to tune the redox activity, with multiple quinones, imides, imines and conductive polymers already explored\cite{Gupta_AEM_15_2025, Fan_PMS_149_2025, Zhang_ESM_2024, Cui_SM_3_2022, Li_AFM_34_2024, Espinoza_FB_2025, Cui_CSC_13_2020, Baker_JES_150_2026, Espinoza_JES_171_2026}. However, organic materials tend to display overall low operating voltages and specific capacities, as well as suffer from structural dissolution into the aqueous electrolyte during cycling, which has severely limited the applicability of organic cathodes in \ac{RAZIB} systems\cite{Cui_SM_3_2022, Cui_CSC_13_2020}. On the other hand, promising battery metrics (\textit{e.g.}, specific capacity, operating voltage, and energy density) have been reported for inorganic \ac{RAZIB} cathode materials\cite{Fan_PMS_149_2025}. Oxides were the first class of materials investigated as cathode materials for \ac{RAZIB}, as \citet{Yamamoto_ICA_117_1986} reported the development of a Zn/\ce{MnO2} battery cell with a mildly acidic \ce{ZnSO4} aqueous electrolyte\cite{Shoji_JAE_18_1988}. Since then, \ce{MnO2} polymorphs and other \ce{Mn} oxides have been extensively explored as cathodes for \ac{RAZIB}, with the eco-friendly \ce{Mn}-based cathode cells reporting both high operating potential (ca. 1.5~V vs Zn/\ce{Zn^{2+}}) and discharge capacity ($>$~200~mA~h~g$^{-1}$)\cite{Liao_ESM_44_2022,  Miliante_JPCC_128_2024, Rubel_JPCC_126_2022}. Current development of oxide materials for \acp{RAZIB} is mainly focused \ce{Mn}- and \ce{V}-based oxides, with only a few studies reporting the investigation of other transition metal centres (\textit{e.g.}, \ce{Mo}, \ce{Cu}) despite the wide range of experimentally realized oxide materials in the literature\cite{Miliante_JPCC_128_2024, Li_AFM_34_2024, He_EA_465_2023, Wu_JAC_895_2022, Gourley_BS_9_2026}.

Polyanionic materials, such as metal phosphates, have also been previously investigated as cathode materials for \acp{RAZIB}, with the scientific interest behind these materials being due to the higher structural stability conferred by the covalent bonds in polyanions\cite{Bin_AFM_33_2023, Li_AFM_34_2024}. The vast majority of polyanionic cathode materials reported in the literature for \ac{RAZIB} are phosphorus-based, as silicon- and sulphur-based polyanions have demonstrated poor reversible intercalation within the water stability potential window and low electrochemical stability in the aqueous electrolyte\cite{Bin_AFM_33_2023}. Materials that follow the olivine (\textit{e.g.}, \ce{LiFePO4}) and \acsu{NASICON} (\acl{NASICON}, \textit{e.g.}, \ce{Na3V2(PO4)3}) structural arrangements comprise the majority of polyanion materials investigated as cathodes for \ac{RAZIB}\cite{Bin_AFM_33_2023}. However, the development of polyanionic cathode materials is still arguably rooted in structures previously investigated for sodium- and lithium-ion batteries, despite the major difference in electrolyte chemistry being found in \acp{RAZIB}\cite{Bin_AFM_33_2023, Ou_CSC_15_2022}. Polyanions are very attractive as \ac{RAZIB} cathodes for their high operating potential (up to  1.9~V vs Zn/\ce{Zn^{2+}}), which can be achieved due to the weaker ionic bonding between the metal centre and the oxygen atom induced by, for example, the \ce{P}-\ce{O} bonds in phosphates \cite{Bin_AFM_33_2023, Li_AFM_34_2024, Ou_CSC_15_2022, Ni_AS_4_2017}. Fluorine is commonly incorporated in polynanions to enhance the inductive effect and increase the \ac{RAZIB} operating voltage, yet its inclusion increases the molecular weight of the active material, which contributes to a lower specific capacity\cite{Bin_AFM_33_2023, Ou_CSC_15_2022}. Chalcogenides and \ac{PBA} have also been previously explored as cathode materials for \ac{RAZIB}, having displayed reasonable battery operating potentials ($>$~1.0~V vs Zn/\ce{Zn^{2+}})\cite{Li_AFM_34_2024, Zampardi_COE_21_2020, Liu_JMCA_12_2024, Wang_JPD_57_2024}. However, the low specific capacity and/or structural instability of chalcogenides and \ac{PBA} cathode materials severely hamper their utilization in commercial \ac{RAZIB} systems\cite{Zampardi_COE_21_2020, Zhou_JCIS_605_2022, Li_AFM_34_2024, Wang_JPD_57_2024}. The stability and specific capacity shortcomings of currently reported cathode materials for \ac{RAZIB} make the proposal of novel materials instrumental in order to fully realize the commercial potential of \acp{RAZIB}.

Theoretical calculations have been regularly employed for the discovery and analysis of cathode materials for rechargeable batteries\cite{Miliante_JPCC_128_2024, Rubel_JPCC_126_2022, Park_EES_14_2021, Mueller_CM_23_2011, Hautier_JMC_21_2011, Kim_CM_37_2024, Zhu_APLM_9_2021}. \citet{Chen_CM_24_2012}, for example, performed high-throughput screening of novel cathode materials for \ac{LIB} from \ac{DFT} calculations of the material stability, lithium intercalation potential and ion migration barrier. The theoretical results were backed up by the experimental validation of the proposed materials crystal structure and electrochemical performance, resulting in the proposal of carbonophosphates as a novel material class for \ac{LIB} cathodes\cite{Chen_CM_24_2012}. \citet{Nishijima_NC_5_2014} systematically investigated cation substitutions in the crystal structure of the well-established \ce{LiFePO4} \ac{LIB} cathode material in order to extend the battery capacity retention. The \ce{Zr}- and \ce{Si} substituted material proposed from the theoretical calculations reported an approximately 6 times improvement in the number of battery cycles to achieve 80$\%$ capacity retention when compared to the pristine material results\cite{Nishijima_NC_5_2014}. To date, only a few studies have taken a theory-based approach for the discovery of novel \ac{RAZIB} cathode materials. For example, \citet{Zhou_ATS_4_2021} trained a machine learning model on the theoretical properties of more than 40,000 inorganic materials available in a database to calculate the respective \ce{Zn^{2+}} intercalation potential for each material, and proposed 8 promising previously unexplored \ac{RAZIB} cathode materials. \citet{Cai_ESM_42_2021} combined high-throughput \textit{ab initio} calculations of relevant cathode metrics, such as ionic conductivity and volume expansion, and machine learning to screen \ce{Zn}-containing spinel structures as \ac{RAZIB} cathode materials. \citet{Wudil_AAMI_17_2025} trained a machine learning model on \ce{Zn}-containing structures reported on the Materials Project database\cite{Jain_AM_1_2013, Ong_CMS_97_2015} to predict both the voltage and capacity of novel \ac{RAZIB} cathode materials. Finally, \citet{Pascazio_CM_37_2025} factored in electrochemical stability effects and \ce{Zn^{2+}} diffusion for the proposal of novel cathode materials for \acp{RAZIB} with high zinc-ion intercalation potentials.

Present studies have provided great scientific insight on the influence of atomic-level cathode properties on the macroscopic \ac{RAZIB} performance, while also leveraging the utilization of machine-learning frameworks and atomic modelling for the discovery of novel cathodes for \acp{RAZIB}. However, there are still open opportunities to improve theoretical cathode screening for the realization of materials with suitable attributes for experimental \ac{RAZIB} testing. For example, the electrochemical stability of the materials in an aqueous environment is widely overlooked during the screening process, putting into question the stability of the proposed materials for prolonged operation in aqueous media. Many candidate cathode materials also reported reversible potentials for \ce{Zn^{2+}} intercalation considerably greater than the \ac{OER} reversible potential, which ultimately would cause the degradation of the aqueous electrolyte employed in \acp{RAZIB}. Also, theoretical-only materials (\textit{i.e.}, materials without an established synthetic route) were proposed for experimental testing in \ac{RAZIB} cells, which casts doubt on the real-world application of the reported theoretical material discovery frameworks. Therefore, taking into account the limited amount of theoretical explorations of  \ac{RAZIB} cathode materials reported in the literature and the persistent overlook of critical factors for appropriate cathode performance in present studies, it is possible to identify the currently underdeveloped potential of computational discovery of novel \ac{RAZIB} cathode materials.

In this study, more than 2000 previously synthesized oxide, chalcogenide, \ac{PBA}, and polyanion materials with structural data reported in open databases were screened for the identification of novel cathode materials for \acp{RAZIB}. The structural availability for \ce{Zn^{2+}} (de)intercalation was evaluated for each material through the calculation of the available \ce{Zn^{2+}} percolation paths in the crystal structure. Electrochemically stable materials for continuous operation as cathodes in an aqueous environment were identified through the calculation of the electrochemical decomposition energy in the \ac{RAZIB} cathode potential and pH window from computationally obtained Pourbaix diagrams. The redox-active metal centres oxidation states before and after the \ce{Zn^{2+}} (de)intercalation were also examined with respect to their experimentally determined oxidation states in order to verify the chemical viability of the proposed material operation as a cathode. Overall, 130 materials were identified with having viable crystal structure, electrochemical stability and chemical environment for operation as active materials in the cathode of \acp{RAZIB}, for which the \acl{EZn} was calculated through molecular dynamic simulations. The oxidation state of the cathode transition metal prior to \ce{Zn^{2+}} intercalation and the \ce{Zn^{2+}} coordination environment after intercalation were revealed as the main parameters governing the calculated magnitude of the reversible \acl{EZn}, which provided significant insight for the development of cathode materials with high battery cycling potentials. \textit{Ab initio} calculations were then combined with \ac{CV} and \ac{GCD} experiments of $\alpha$-\ce{FePO4} as a cathode material for \acp{RAZIB}, which identified this material as a false positive of the screening and highlighted the limits of the descriptors considered. Finally, three novel materials in \ce{CaFe5P5O22}, \ce{RbV(PO4)2}, and \ce{Mo2P2O11} were proposed for experimental investigation for the first time due to their higher operating potential, electrochemical stability in aqueous media, gravimetric capacity and energy density, directly supporting the development of \acp{RAZIB} as commercially viable battery technology to aid in the ongoing energy transition.

\section{Methods}

The structures to be explored as potential \ac{RAZIB} cathode materials were retrieved from the Materials Project database  (version \textit{2025.09.25}) via its \acsu{API} (\acl{API}) \cite{Ong_CMS_97_2015, Jain_AM_1_2013}. Four classes of materials were considered for the investigation based on their previous utilization as \ac{RAZIB} cathode materials: oxides, chalcogenides, \acp{PBA}, and polyanions. The database was queried with respect to the elements considered to be forming the chemical systems of each class. The queried chemical systems for each class are presented in Table~\ref{tbl:QuerySystems}, with the elements chosen to be investigated being selected based on their toxicity, price and presence in previously explored cathode materials. For example, to get the data for all \ce{Mn}-based oxides containing \ce{Li} available in the database, a query equal to `\ce{Li}-\ce{Mn}-\ce{O}' would be utilized. A parenthesis in the chemical systems presented in Table~\ref{tbl:QuerySystems} indicates that both systems which include and do not include the element inside the parenthesis (or an element of the group inside the parenthesis in the case of \textit{Alk}) were considered during the data retrieval process from the database. Also, the presence or absence of \ce{Zn} in the structure was considered (`(\ce{Zn})' in the query) in order to completely explore all potential \ac{RAZIB} cathode materials available for all classes, be they initially at a charged (\textit{i.e.}, without \ce{Zn} in the structure) or discharged (\textit{i.e.}, with \ce{Zn} in the structure) battery state. Only oxide materials containing alkaline and alkaline-earth metals were considered here, since a thorough exploration of binary oxides (corresponding query of `\textit{Mtl}-\ce{O}-(\ce{Zn})') as cathode materials for \ac{RAZIB} was already performed by our group \cite{Miliante_JPCC_128_2024}.  All possible combinations that can be formed from the elements in each query were considered, resulting in 2570 unique chemical systems: 180 for oxides,  600 for chalcogenides, 990 for \acp{PBA}, and 800 for polyanions. In order to guarantee that materials explored in this study can be directly experimentally investigated as \ac{RAZIB} cathode materials, only structures that were reported by the database as previously synthesized were considered during the query and data retrieval process.

\newcolumntype{K}[1]{>{\centering\arraybackslash}m{#1}}
\setlength\extrarowheight{0pt}
\setlength{\tabcolsep}{6.25pt}
\begin{table}
  \caption{Queried systems from the Materials Project database\cite{Ong_CMS_97_2015, Jain_AM_1_2013} with respect to the elements present in their formulas, where \textit{Alk} indicates an alkali or alkaline earth metal (\ce{Li}, \ce{Na}, \ce{K}, \ce{Rb}, \ce{Cs}, \ce{Mg}, \ce{Ca}, \ce{Sr}, or \ce{Ba}), \textit{Mtl} is a transition metal (\ce{Ti}, \ce{V}, \ce{Cr}, \ce{Mn}, \ce{Fe}, \ce{Co}, \ce{Ni}, \ce{Cu}, \ce{Mo}, or \ce{W}), and \textit{Chl} is a chalcogen (\ce{S}, \ce{Se}, or \ce{Te}). \textit{Mtl$_{1}$} and \textit{Mtl$_{2}$} are two different transition metal species considered during the query process.}
  \label{tbl:QuerySystems}
  
    \centering
    \begin{tabular}{cc}
    \hline
    Class & Query \\ \hline
    Oxides & \textit{Alk}-\textit{Mtl}-\ce{O}-(\ce{Zn}) \\ 
    Chalcogenides & (\textit{Alk})-\textit{Mtl}-\textit{Chl}-(\ce{Zn}) \\ 
    PBAs & (\textit{Alk})-\textit{Mtl$_{1}$}-\textit{Mtl$_{2}$}-\ce{C}-\ce{N}-(\ce{Zn}) \\ 
    Polyanions & \textit{Alk}-(\ce{F})-(\ce{H})-\textit{Mtl}-\ce{O}-\ce{P}-(\ce{Zn}) \\ \hline
    \end{tabular}
\end{table}

\begin{figure}
\includegraphics{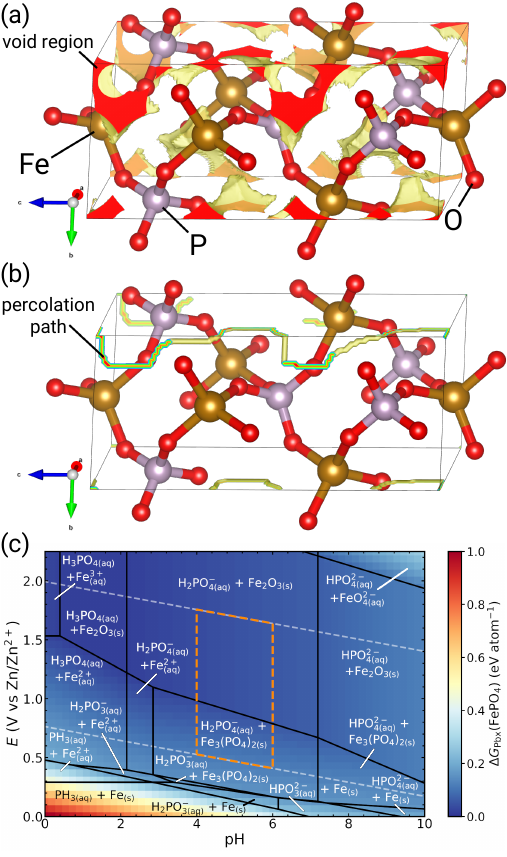}
\caption{(a) Result for the void search in the $\alpha$-\ce{FePO4} structure, with the regions in the crystal which are at least 1.90~$\text{\AA}$ from an atom being presented. (b) Percolation path associated with the \acf{percdistmax} for $\alpha$-\ce{FePO4}, calculated to be equal to 1.87~$\text{\AA}$ with an associated \acf{percdisp} of 0.99~$\text{\AA}$. (c) Calculated Pourbaix diagram for the \ce{Fe}-\ce{P}-\ce{H2O} system, alongside a heatmap highlighting the calculated \acf{dGpbx} for $\alpha$-\ce{FePO4}. The region of interest for \ac{RAZIB} operation is delimited by the orange dashed lines.}
\label{fig:MethodExample}
\end{figure}

For a material to be a possible \ac{RAZIB} cathode, it is necessary for the material to have sufficient space/voids in its crystal structure for the reversible \ce{Zn^{2+}} (de)intercalation, with the ample space for ionic intercalation potentially granting the active material higher structural stability during cycling. For this reason, a void search algorithm was coded to parse the crystal structure of all investigated materials and determine those which have promising structures for cathode operation. The void search algorithm works as follows. First, the crystal lattice parameters and the atomic positions in the lattice are read from the data retrieved from the database for the material in question. Then, an evenly spaced 3D grid is built based on the cell lattice parameters, with the distance between each point in the grid and each atom in the structure (and its periodic images) being then calculated. The smallest distance to an atom in the structure is then determined for each point in the grid, with the distance data being vital in the analysis of the crystal structure, as it allows for the numeric quantification of how ``empty" (\textit{i.e.}, filled with voids) the structure is depending on the considered distance threshold to an atom in the lattice. For example, Figure~\ref{fig:MethodExample}a shows all the positions inside the $\alpha$-\ce{FePO4} (mp-19109) lattice which are at least 1.90~$\text{\AA}$ from an atom in the cell, allowing for the identification of regions in the cell for which \ce{Zn^{2+}} ions could establish themselves in and bond to the atoms in the host structure after intercalation. A maximum spacing between the 3D grid points of 0.01~$\text{\AA}$ was chosen for the calculations performed in this study. Also, it is important to note that the chemical nature of the atomic elements is not taken into account when calculating the distance from the grid points to the atoms; therefore, the impact of the atomic radius inside the cell, for example, is not relevant in the void space calculation.

The calculation of the void spaces allows for the determination of the materials with suitable ``empty gaps" in their structures for \ce{Zn^{2+}} intercalation. However, it is important for the intercalating \ce{Zn^{2+}} ion to have not only space to coordinate with the host structure but also an unobstructed path in the structure for it to percolate through during the battery operation. It is then necessary to calculate the available  \ce{Zn^{2+}} percolation paths, and their complexity, in order to determine materials which have a promising crystal structure to operate as a \ac{RAZIB} cathode. For example, if a \ce{Zn^{2+}} ion percolates to close to other atoms in the structure, the energy barrier associated with the movement of the \ce{Zn^{2+}} ion will be considerably high and consequently deem the ion percolation unfeasible to occur. Also, if the percolation path is very elaborate/diverge too much from what would be a straight percolation path, it will be impractical for the \ce{Zn^{2+}} to percolate inside the structure. Therefore, to account for an unimpeded \ce{Zn^{2+}} percolation inside the structure and the viability of a percolation path, two parameters are introduced here. The first parameter is the \ac{percdist} and it captures the minimum distance that all points along an evaluated percolation path are from an atom in the structure. Materials with higher \ac{percdist} will then have a more unobstructed percolation path for \ce{Zn^{2+}} movement (\textit{i.e.}, unimpeded \ce{Zn^{2+}} percolation). The second parameter is the \ac{percdisp}, which is the maximum distance that a percolation path deviates from being a straight path inside the crystallographic cell. Thus, the closer to 0~$\text{\AA}$ (value for straight percolation path) the calculated \ac{percdisp} is the more direct the \ce{Zn^{2+}} percolation within the crystal cell will be.

A percolation path can be calculated based on the 3D grid points established for the void space results, since they already capture the distance to the atoms for all positions inside the lattice, a necessary information for determining \ac{percdist}. A percolation path can be theoretically considered to start at any point inside the lattice (\textit{e.g.}, $r_{\text{i}}$~=~(0.5, 0.5, 0.5) in fractional coordinates), however, it is necessary for the path to end at the same lattice point in one of the periodic images (\textit{e.g.},$r_{\text{f}}$~=~(1.5, 0.5, 0.5)). This requirement guarantees that the entire structure would be crossed by the percolating ion if it followed the calculated path, as it would return to a position crystallographically identical to its starting point. In the algorithm established here for the calculation of the percolation path, first, a high initial \ac{percdist} guess value is chosen (15~$\text{\AA}$ was used here). Then, the shortest path that can be established by connecting adjacent grid points that are at least \ac{percdist} from an atom in the structure is calculated, taking into account the discussed requirement for the start and end point. If the percolation path could not be established, then the evaluated \ac{percdist} is decreased by a small factor (0.01~$\text{\AA}$ was used here) and the percolation path search is restarted. However, if a percolation path is found, then a viable \ac{percdist} value for the structure is found. The \ac{percdist} found through this method is also then called the \ac{percdistmax}, since it is the highest distance for which a percolation path can be established inside the crystal cell. For any distance of less than \ac{percdistmax}, it will be possible to establish a percolation path inside the structure, with these distinct percolation paths also having different \ac{percdist} and \ac{percdisp} values associated with them. The percolation path for $\alpha$-\ce{FePO4} with a \ac{percdistmax} equal to 1.87~$\text{\AA}$ and an associated \ac{percdisp} of 0.99~$\text{\AA}$ is shown as an example in Figure~\ref{fig:MethodExample}b.

The utilization of aqueous electrolytes in \ac{RAZIB} adds further stability considerations when evaluating the proposal of novel cathode materials, as active material dissolution into the electrolyte has been shown to occur in both organic and inorganic \ac{RAZIB} cathode materials during cycling\cite{Cui_SM_3_2022, Wu_EES_13_2020, Zhou_JCIS_605_2022}. The electrochemical stability in an aqueous environment at \ac{RAZIB} operating conditions was evaluated through an analysis of computationally obtained Pourbaix diagrams, an approach which has already been shown to appropriately capture the stability limitation of cathode materials in aqueous batteries\cite{Miliante_JPCC_128_2024, Rubel_JPCC_126_2022, Bischoff_JES_167_2020}. The diagrams for the electrochemical stability analysis were constructed with the use of the Materials Project Pourbaix diagram tool\cite{Persson_PRB_85_2012, Singh_CM_29_2017, Patel_PCCP_21_2019}, which reports the \acf{dGpbx} for a material at  different applied potentials and pH conditions evaluated during diagram construction. The \ac{dGpbx} shows how unstable an investigated material is with respect to the thermodynamically stable material at the potential and pH condition of interest. For this reason, materials with lower \ac{dGpbx} values can be considered to be more electrochemically stable for use as a \ac{RAZIB} cathode material. The conditions of interest for the electrochemical stability analysis were an \ac{elect_pot} between the reversible potentials for the \acf{HER} and \acf{OER}, to avoid the degradation of the aqueous electrolyte, and pH levels between 4 and 6, considering previous reports on electrolyte pH fluctuation during \ac{RAZIB} operation\cite{Bischoff_JES_167_2020, Rodriguez_JMCA_9_2021}. The \ac{avgdGpbx} is then calculated with respect to the \ac{dGpbx} values reported within the \ac{elect_pot}-pH operation window to quantify the electrochemical instability of the materials under the conditions of interest for \ac{RAZIB} cathode operation. The calculated Pourbaix diagram for $\alpha$-\ce{FePO4} (\ac{avgdGpbx}~=~0.06~eV~atom$^{-1}$) is presented in Figure~\ref{fig:MethodExample}c as an example, with a heatmap displaying the calculated \ac{dGpbx} values for the material at different \ac{elect_pot} and pH conditions.

The transition metal atoms undergo a redox reaction during the \ce{Zn^{2+}} (de)intercalation process. Therefore, it is important to verify that the oxidation states attained by the metal centres after the (de)intercalation continue to be chemically feasible. The experimentally observed oxidation states for all transition metals considered in this study were first compiled, with the theoretical metal centre oxidation states in each material prior to (de)intercalation also calculated. For materials that are retrieved from the database at an expected charged battery state (\textit{i.e.}, without \ce{Zn} in the structure), if the metal centre in the structure was already at the lowest experimental oxidation state for that species, the material was then considered invalid for use as a cathode material. The material disqualification is due to the chemical inaccuracy that would be achieved, since the evaluated metal centre would be predicted to reduce to an oxidation state lower than what has been experimentally verified after \ce{Zn^{2+}} ions intercalate into the structure. Conversely, for materials that are reported to be at a discharged state (\textit{i.e.}, materials with \ce{Zn} in the structure), the structure is considered invalid for use as cathode if the metal centre already is on its highest oxidation state prior to \ce{Zn^{2+}} deintercalation. All other oxidation states for the transition metals were considered valid.

It is then necessary to determine if the \ce{Zn^{2+}} (de)intercalation process is thermodynamically viable to occur in the studied structures to validate their application as cathodes for \ac{RAZIB}. The energetic feasibility for \ce{Zn^{2+}} (de)intercalation can be probed by calculating the \ac{HZn} (in eV) 
\begin{equation}\label{eqn:HZn}
  \Delta H_{\mathrm{Zn}} = U_{\mathrm{tot}}(\mathrm{M_{dchg}}) - U_{\mathrm{tot}}(\mathrm{M_{chg}}) - zu_{\mathrm{tot}}(\ce{Zn}_{\mathrm{bulk}}),
\end{equation}
where $U_{\mathrm{tot}}(i)$ is the total energy (in eV) calculated for system $i$, \acs{mdchg} and \acs{mchg} are the studied materials at a discharge and charged battery state, respectively, $z$ is the number of \ce{Zn} atoms that (de)intercalated, $u_{\mathrm{tot}}(i)$ is the total energy per atom (in eV~atom$^{-1}$) calculated for system $i$, and \ce{Zn}$_{\mathrm{bulk}}$ is the bulk structure of \ce{Zn}. A negative \ac{HZn} value indicates that the \ce{Zn^{2+}} (de)intercalation process is energetically viable to occur in the structure, while positive values indicate that the process is energetically impractical to occur. The \ac{EZn} (in V~vs.~\ce{Zn}/\ce{Zn^{2+}}) can then be calculated based on the Nerst equation
\begin{equation}\label{eqn:EZn}
  E_{\mathrm{Zn}} = -\frac{\Delta H_\mathrm{Zn}}{\upsilon e},
\end{equation}
where $\upsilon$ is the number of electrons transferred in the \ce{Zn^{2+}} (de)intercalation process and $e$ is the elementary charge. 

The structure \acl{Etot} was calculated from \ac{MD} simulations run with the MACE mp-0 interatomic potential\cite{Batatia_ANIPC_35_2022, Batatia_2023}, which has been parametrized with respect to the \ac{DFT} data available in the Materials Project database\cite{Jain_AM_1_2013, Ong_CMS_97_2015}. The molecular dynamics simulations were run for 50~ps with a 1~fs timestep under a NPT ensemble following the methodology proposed by \citet{Melchionna_MP_78_1993}\cite{Melchionna_PRE_61_2000}. Constant Nosé-Hoover thermostat (\textit{ttime}~=~25~fs) and Parinello-Rahman barostat (\textit{pfactor}~=~$10^{6}$~GPa~fs$^{2}$) respectively set to 300~K and 1~bar were employed, with all \ac{MD} simulations being run utilizing the \ac{ASE} Python library\cite{Larsen_JPCM_29_2017}.

To evaluate the feasibility of the candidate materials as cathodes for \acp{RAZIB}, we calculated the initial theoretical potential during battery discharge (or charge) through the incorporation (or removal) of a single \ce{Zn} atom into (from) the cathode. This parameter is called the \ac{EZninitial}, and it is directly applicable to the screening of cathode materials, as it captures the predicted first potential to be reported by the battery during cycling, allowing for the direct determination of whether the material should continue to be considered during the screening process from its initial \ac{EZn} reported at the beginning of the cycle. To create the \acs{mdchg} and \acs{mchg} supercells for the calculation of \ac{EZninitial}, the initial crystallographic cells obtained from the Materials Project database were replicated in all 3 dimensions, so at least 100 atoms were present in the supercells. If the phase extracted from the database already contained \ce{Zn} in its structure (\acs{mdchg} - \textit{e.g.},  \ce{SrV2ZnO7}, \ce{NaVZnP2O9}), the \acs{mchg} phase was created by removing a random \ce{Zn} atom present in the \acs{mdchg} supercell (\textit{i.e.}, \ce{Zn^{2+}} deintercalation process). However, if the material obtained from the database did not contain \ce{Zn} (\acs{mchg} - \textit{e.g.}, \ce{LiMoP2O7}, \ce{K2Ti6O13}), the initial \acs{mdchg} phase was then created by placing a \ce{Zn} atom on the position most distant from other atoms on the calculated \ac{percdistmax} percolation path for the associated \acs{mchg} structure (\textit{i.e.}, \ce{Zn^{2+}} intercalation process). A value of $z$ of 1 and $\upsilon$ of 2 was then considered for the calculation of \ac{EZninitial} (see Eqs.~\eqref{eqn:HZn}~and~\eqref{eqn:EZn}), since there was only one divalent \ce{Zn^{2+}} ion difference between the atomic compositions of the \acs{mdchg} and \acs{mchg} structures. 

The \ac{EZnavg} associated with the variation of one integer oxidation state at the transition metal sites during a battery cycle (\textit{e.g.}, \textit{Mtl}$^{\text{III}}_{\text{(s)}}$ to \textit{Mtl}$^{\text{II}}_{\text{(s)}}$) was also calculated for the most promising materials to probe the change in \ac{EZn} throughout the discharge cycle when compared to the calculated \ac{EZninitial}. The most promising materials for implementation in \acp{RAZIB} will then have both high \ac{EZninitial} and \ac{EZnavg}, as this will demonstrate that the associated \ce{Zn^{2+}} intercalation potential is able to remain elevated throughout the battery cycle. The \acs{mdchg} structures for the calculation of \ac{EZnavg} were created by randomly incorporating \ce{Zn} atoms into the identified percolation paths of the \acs{mchg} structures until a 1:2 ratio of divalent \ce{Zn^{2+}} to \textit{Mtl} was achieved. The respective amount of \ce{Zn} incorporated into \acs{mchg} was then considered in the calculation of \ac{HZn} and \ac{EZnavg}, as described in Eqs.~\eqref{eqn:HZn} and \eqref{eqn:EZn}, with the presented \ac{EZnavg} being calculated as the average of the \ac{EZnavg} results from 5 random \ce{Zn^{2+}} placement configurations.

Relevant metrics for \ac{RAZIB} operation, such as \ac{Qw} and \ac{Ww}, were calculated for the proposed materials for experimental investigation. The \ac{Qw} (in mA~h~g$^{-1}$) for a cathode material can be calculated from
\begin{equation}\label{eqn:Qw}
  Q_\text{w} = \frac{n_{\text{e}} \text{F}}{3.6w},
\end{equation}
where $n_{\text{e}}$ is the number of electrons considered in the battery discharge reaction per cathode formula, F is the Faraday constant (96,485~C~mol$^{-1}$), and $w$ is the molecular weight of the cathode material at its charged state (in g~mol$^{-1}$). The variation of one integer amount in the oxidation state of the transition metal sites was also considered here for the calculation of \ac{Qw} (\textit{e.g.}, \textit{Mtl}$^{\text{III}}_{\text{(s)}}$ to \textit{Mtl}$^{\text{II}}_{\text{(s)}}$), thus $n_{\text{e}}$ was equal to 1 for all materials. The \ac{Ww} (in W~h~kg$^{-1}$) can then be directly calculated from the \ac{EZnavg} and \ac{Qw} results
\begin{equation}\label{eqn:Ww}
  W_\text{w} = Q_\text{w}\langle E_{\text{Zn}}\rangle.
\end{equation}

\Acf{DFT} calculations for the \textit{olivine}-, $\alpha$-, and $\beta$-\ce{FePO4} polymorphs were carried out with projector augmented wave pseudopotentials~\cite{Kresse_PRB_59_1999} (14, 5, and 6 valence electrons considered for \ce{Fe}, \ce{P}, and \ce{O}, respectively), utilizing the \acl{PBE} exchange-correlational functional~\cite{Perdew_PRL_77_1996} as implemented in \acs{VASP}~\cite{Kresse_PRB_47_1993, Kresse_CMS_6_1996, Kresse_PRB_54_1996} (version 6.4.3). A plane-wave cutoff energy of 520~eV was utilized in all calculations, with the DFT-D3 method with Becke-Johnson damping function employed to account for the van der Waals interactions in the materials\cite{Grimme_JCP_132_2010, Grimme_JCC_32_2011, Becke_JCP_123_2005, Johnson_JCP_123_2005, Johnson_JCP_124_2006}. Initial ferromagnetic ordering for spin-polarized calculations was employed for all materials, following the reported magnetism for the investigated materials from the database\cite{Ong_CMS_97_2015, Jain_AM_1_2013}. Hubbard \textit{U} correction in the formulation of \citet{Dudarev_PRB_57_1998} was applied to the $d$ electrons of \ce{Fe} (\textit{U}$_{\text{eff}}$~=~5.3~eV\cite{Jain_PRB_84_2011, Jain_AM_1_2013}). A $\Gamma$-centred \textit{k}-point grid of 6$\times$4$\times$2, 6$\times$6$\times$3, and 6$\times$6$\times$3 were respectively employed for the calculations for \textit{olivine}-\ce{FePO4} (mp-20361), $\alpha$-\ce{FePO4} (mp-19109), and $\beta$-\ce{FePO4} (mp-19752). The structures obtained from the database were submitted to cell and atomic position relaxation using a Gaussian smearing of 0.05~eV, with convergence tolerances of 10$^{-7}$~eV and 10$^{-1}$~eV~$\text{\AA}^{-1}$, respectively, considered for the energy change in consecutive self-consistent loops and for the ionic forces on each atom.

The relaxed structures were then used in a single static calculation with the same settings, except that the Brillouin zone integration was performed with the tetrahedron method with Bl\"ochl corrections\cite{Blochl_PRB_49_1994}, symmetry was switched off so that the \textit{k}-point grid was reduced by time-reversal symmetry only, as required by LOBSTER, and the number of bands was increased to exceed the number of local basis functions used in the projection.

The bonding interactions between \ce{Fe} and \ce{O} atoms were evaluated via the \ac{pCOHP} computed with the LOBSTER package (version 6.0.0)\cite{Dronskowski_JPC_97_1993, Deringer_JPC_115_2011, Maintz_JCC_34_2013, Maintz_JPC_37_2016}, which projects the plane-wave wavefunctions onto a local atomic-orbital basis. The pbeVaspFit2015 basis set\cite{Maintz_JPC_37_2016} was used with the \ce{Fe} 3$p$, 3$d$ and 4$s$, \ce{P} 3$s$ and 3$p$, and \ce{O} 2$s$ and 2$p$ orbitals, and the \ac{pCOHP} was computed for all \ce{Fe}-\ce{O} pairs with interatomic distances between 1.6 and 2.5~$\text{\AA}$ in an energy window from $-$15 to 10~eV relative to the \ac{Ef}. The \ac{-IpCOHP} values reported here were obtained by integrating the \ac{pCOHP} of each \ce{Fe}-\ce{O} pair up to the \ac{Ef}, as listed by LOBSTER, and are given per bond or summed over the coordination shell of each \ce{Fe} atom, as indicated in the text. The absolute charge spilling was below 1.5\% for both spin channels in \textit{olivine}- and $\alpha$-\ce{FePO4}.

The code written for this project (void search, percolation path and electrochemical stability analysis) was programmed in Python and is hosted on \href{https://github.com/CMMiliante/ABCs}{GitHub repository}.

\section{Results and discussion}

In total, the data for 2044 previously synthesized materials were extracted from the Materials Project database\cite{Jain_AM_1_2013, Ong_CMS_97_2015} based on the queried chemical systems, with 719 materials being from the oxide class, 509 chalcogenides, 27 \acp{PBA}, and 789 polyanions. \ce{Fe} (365 materials), \ce{V} (305 materials) and \ce{Mn} (245 materials) were the transition metals with the highest frequency in the materials obtained from the database, comprising together approximately 45~$\%$ of all materials investigated. A summary of all materials reported from the database with respect to their classes and transition metal centres is shown in Figure~S1. 

\subsection{Electrochemical stability screening}

\begin{figure}
\includegraphics{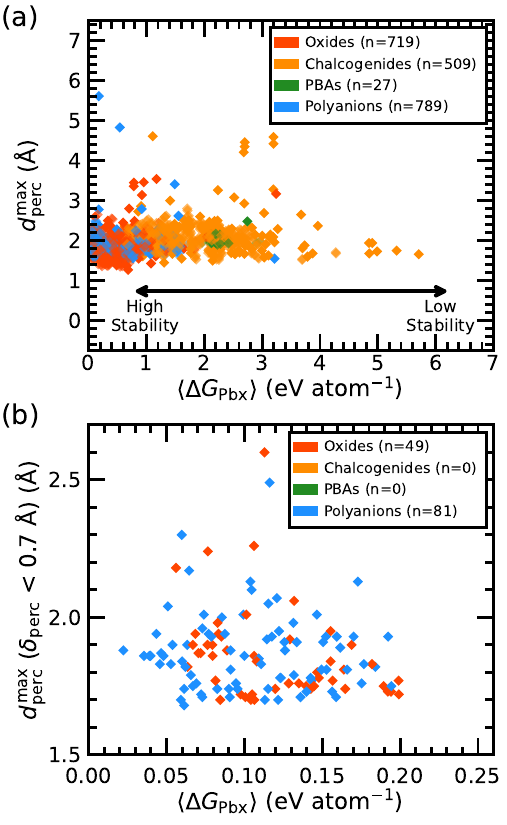}
\caption{(a) \Acf{percdistmax} as a function of the  \acf{avgdGpbx} for all materials considered. (b) \Ac{percdistmax} for a \acf{percdisp} of less than 0.7~$\text{\AA}$ (\ac{percdistmax}(\ac{percdisp}~$<$~0.7~$\text{\AA}$)) and \ac{avgdGpbx} results for materials considered for \ac{EZn} calculation.}
\label{fig:ElectStabxPercPath}
\end{figure}

The electrochemical stability of the host material was evaluated through the calculation of the \acf{avgdGpbx} (see Methods section). \ac{avgdGpbx} captures the average degree of energy instability for an investigated material with regard to the most stable phase(s) for the chemical system within the \ac{elect_pot} and pH window considered for cathode cycling in \acp{RAZIB}. A \ac{avgdGpbx}($\zeta$) equal to 0~eV~atom$^{-1}$ would indicate that the system $\zeta$ is the most stable phase for the chemical system at the conditions in question, with low active cathode material degradation being expected for $\zeta$ during cycling. Therefore, the lower the value for \ac{avgdGpbx}, the higher the electrochemical stability in aqueous media for the material is, as it will be closer in energy to the most stable phase at the considered \ac{elect_pot} and pH window. Lower electrochemical stability is found for materials with higher \ac{avgdGpbx} results. The \ac{avgdGpbx} results for all considered materials are presented on the horizontal axis of the graph in Figure~\ref{fig:ElectStabxPercPath}a, with the results separated per material class being shown in Figure~S2. The compiled \ac{avgdGpbx} results present a clear relationship between electrochemical stability and material class. Oxides and polyanions are concentrated at lower \ac{avgdGpbx} values, with only a few materials from these classes reporting \ac{avgdGpbx} greater than 1~eV~atom$^{-1}$ (Fig.~\ref{fig:ElectStabxPercPath}a~and~S2). On the other hand, chalcogenides and \acp{PBA} reported higher \ac{avgdGpbx}, with a considerable range of \ac{avgdGpbx} results being captured for chalcogenides (approximately from 0.5 to 3.3~eV~atom$^{-1}$). The electrochemical stability of each class was then individually investigated in order to understand the relationship between material class and \ac{avgdGpbx} for the proposal of novel \ac{RAZIB} cathodes.

To understand the wide range of results for the electrochemical stability of chalcogenides, the calculated \ac{avgdGpbx} values for the class were plotted with respect to the individual chalcogen atom present in the materials (\textit{i.e.}, \ce{S}, \ce{Se}, or \ce{Te}), as shown in Figure~S3. It is possible to attest that there is a clear relationship between \ac{avgdGpbx} and the chalcogen atom centre, with \ce{Se}-containing materials being more electrochemically stable, followed by materials with \ce{Te} and then \ce{S}. This trend between the chalcogen element and calculated \ac{avgdGpbx} can be explained by the electrochemical stability of each element in aqueous solutions as captured by their individual experimentally obtained Pourbaix diagram\cite{Pourbaix1974}. In the conditions of interest for \acp{RAZIB} (\textit{i.e.}, $\acs{elect_pot}_{\text{HER}}$ $<$ \ac{EZn} $<$ $\acs{elect_pot}_{\text{OER}}$ and 4 $<$ pH $<$ 6, see Methods section), \ce{S} is thermodynamically favoured to form an aqueous species \ce{SO^{2-}_{4(aq)}}, with solid phases, such as \ce{S_{(s)}} and \ce{H2S_{(s)}}, only being established at potentials close to \ac{HER}\cite[chap.~IV, sec.~19.2]{Pourbaix1974}. On the other hand, solid phases of \ce{Te} (\ce{Te_{(s)}}, \ce{TeO2_{(s)}}\cite[chap.~IV, sec.~19.4]{Pourbaix1974}), and especially \ce{Se} (\ce{Se_{(s)}}\cite[chap.~IV, sec.~19.3]{Pourbaix1974}), were observed to be formed over a wide potential and pH range within the conditions of interest for \acp{RAZIB}. The prevalence of the formation of solid \ce{Se} and \ce{Te} phases from the analysis of their Pourbaix diagram points to an overall higher electrochemical stability (\textit{i.e.}, lower \ac{avgdGpbx}) for solid materials containing these chalcogen elements than those which contain \ce{S}. However, despite the lower \ac{avgdGpbx} results found for chalcogenides containing \ce{Se} and \ce{Te}, chalcogenide materials were still considered to be highly unstable in aqueous media (\textit{i.e.}, high \ac{avgdGpbx}), which will cause chalcogenides to be more prone to degradation and capacity fade when operating as cathode materials for \acp{RAZIB}. This conclusion is in agreement with previous reports on bulk chalcogenide materials as \ac{RAZIB} cathodes, in which modifications to the chalcogenide structure were necessary to be implemented in order to achieve reasonable capacity retention at practical current densities for grid-scale application\cite{Liu_Nanomaterials_19_2022}.

\begin{figure}
\includegraphics{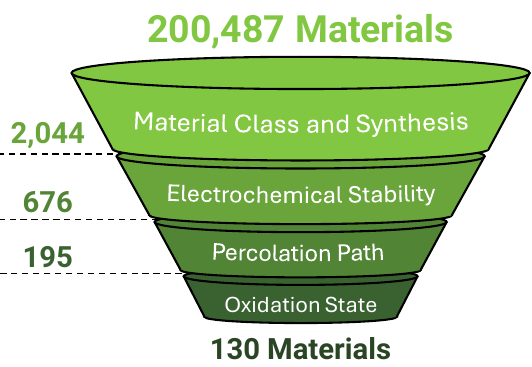}
\caption{Screening funnel for the discovery of novel active materials to be employed in the cathode for \acp{RAZIB}, starting from the 200,487 materials available on the Materials Project database \cite{Jain_AM_1_2013, Ong_CMS_97_2015}. Each section on the funnel indicates a parameter considered during cathode material screening, with the number of remaining candidate materials after each section being also indicated.}
\label{fig:ScreeningFunnel}
\end{figure}

The \ac{PBA} materials investigated also reported high \ac{avgdGpbx} values, with the results concentrated around 2~eV~atom$^{-1}$, indicating an overall low electrochemical stability for \acp{PBA} as \ac{RAZIB} cathodes. The \ac{avgdGpbx} results for the \acp{PBA} can also be explained from the investigation of the Pourbaix diagrams of the constituent elements, similarly to how it was done for the \ac{avgdGpbx} in chalcogenides. From the Pourbaix diagrams of \ce{C} and \ce{N} it is possible to see that the formation of aqueous dissolved species is favoured in the \ac{RAZIB} conditions of interest for both elements (\ce{H2CO_{3(aq)}} for \ce{C}\cite[chap.~IV, sec.~17.1]{Pourbaix1974}, and \ce{NH^{+}_{4(aq)}} and \ce{NO^{-}_{3(aq)}} for \ce{N}\cite[chap.~IV, sec.~18.1]{Pourbaix1974}). The presence of dissolved species of \ce{C} and \ce{N} would then explain the instability of \acp{PBA} in aqueous environments as captured by the high \ac{avgdGpbx} results captured for the class. However, it is important to note that only 27 experimentally obtained materials with \acp{PBA} chemistry were able to be identified from the database, which ultimately limits the conclusions shown here with respect to the electrochemical stability of \ac{PBA} cathodes. In order to screen for materials with high electrochemical stability as cathodes for \acp{RAZIB}, only materials with calculated \ac{avgdGpbx} lower than or equal to 0.2~eV~atom$^{-1}$ continued to be considered during the screening process. The electrochemical stability criteria considerably reduced the number of candidate materials from 2044 to 676, as shown in Figure~\ref{fig:ScreeningFunnel}. 501 of the remaining materials were polyanions and 175 oxides, with all the chalcogenides and \acp{PBA} being filtered out, as can be expected from the \ac{avgdGpbx} discussion.

\subsection{\ce{Zn^{2+}} percolation path analysis}

It is common practice in the literature to evaluate the movement of an intercalating ion in battery cathode materials through the calculation of the percolation energy barrier using, for example, the \ac{NEB} method\cite{Zhao_EER_5_2022, Rong_JCP_145_2016}. Such calculations are able to provide valuable insight into the cathode material by mapping the minimum energy path for ionic percolation in the structure and the associated \ac{Eb}\cite{Zhao_EER_5_2022}. However, the computational expense associated with accurately capturing the optimal path for ion percolation, and its associated energy barrier, is considerably high, since individual \ac{DFT} calculations are necessary for each ionic position image in the studied path. The computational cost for energy barrier calculation made it unfeasible for such calculations to be directly implemented in this study, given the elevated number of materials being investigated. Therefore, a faster and computationally cheaper method for the identification of viable structures for \ce{Zn^{2+}} percolation based on the calculation of \ac{percdistmax} and \ac{percdisp} was implemented. To evaluate the capacity of the \ac{percdistmax} parameter to accurately capture materials with viable \ce{Zn^{2+}} percolation during battery cycling, we calculated the associated \ac{Eb} through \ac{NEB} calculations for a few representative materials spanning a wide range of calculated \ac{percdistmax} (computational details available in the \ac{SI}). The results are presented in Figure~S4, where an inverse relationship between \ac{Eb} and \ac{percdistmax} can be seen, with materials with higher \ac{percdistmax} having a lower calculated \ac{Eb} and vice versa. Therefore, the data displayed in Figure~S4 showcase how the \ac{percdistmax} parameter can be employed as an efficient descriptor to evaluate the viability for \ce{Zn^{2+}} percolation, as it qualitatively captures the \ac{Eb} obtained from \textit{ab initio} calculations at a considerably reduced computational cost.

The ability for \ce{Zn^{2+}} ions to percolate through the structure was then evaluated through an investigation of the available percolation paths inside the crystal structures, with the calculated \ac{percdistmax} for all considered materials being presented on the vertical axis of Figure~\ref{fig:ElectStabxPercPath}a. As discussed previously in the Methods section, the calculated \acf{percdist} is also dependent on the \ac{percdisp}, which accounts for how linear (or not) a given percolation path is. The more linear a percolation path is, the easier it can be expected for the \ce{Zn^{2+}} to traverse through the given path, since there is a more direct path for \ce{Zn^{2+}} intercalation in the material. The effect of \ac{percdisp} on how viable the percolation of \ce{Zn^{2+}} ions is can be attested by comparing the paths for materials with the same calculated \ac{percdistmax}, but different \ac{percdisp}. For example, a \ac{percdistmax} of 1.75~$\text{\AA}$ was calculated for both \ce{Li3VO4} (mp-19219) and \ce{Sr2V2O7} (mp-19660), with their respective paths being shown in Figure~S5. The percolation path found for \ce{Li3VO4} is straight (\ac{percdisp}~=~0.00~$\text{\AA}$) throughout the crystal cell (see Figure~S5a,b), which will cause for the ion percolation to be less challenging, while a more obstructed percolation will be found in \ce{Sr2V2O7} (\ac{percdisp}~=~4.74~$\text{\AA}$) due to the complexity of the path (see Figure~S5c,d). Therefore, it is necessary to take into consideration both \ac{percdistmax} and \ac{percdisp} when screening for potential materials to be employed as \ac{RAZIB} cathodes. For this reason, the \ac{percdistmax} for each material at different maximum  \ac{percdisp} values were also calculated, allowing for a more precise comparison of \ac{percdistmax} values between materials, since all parameters would be under the same displacement criteria.

To determine the \ac{percdistmax} and \ac{percdisp} parameters to be considered in the screening process, the percolation path in commonly employed cathode materials for \ac{RAZIB} was calculated\cite{Fan_PMS_149_2025}. Unfortunately, there is a limited number of materials previously explored as cathodes for \acp{RAZIB} that are also from chemistries not being evaluated in this study, a restriction necessary in order to avoid potential bias in the choice of \ac{percdistmax} and \ac{percdisp} utilized during materials screening. The materials selected were: $\alpha$-\ce{MnO2} (mp-19395) (\ac{percdistmax}~=~2.48~$\text{\AA}$ with \ac{percdisp}~=~0.00~$\text{\AA}$), $\beta$-\ce{MnO2} (mp-510408) (\ac{percdistmax}~=~1.60~$\text{\AA}$ with \ac{percdisp}~=~0.62~$\text{\AA}$), \ce{V2O5} (mp-25279) (\ac{percdistmax}~=~1.80~$\text{\AA}$ with \ac{percdisp}~=~0.00~$\text{\AA}$), \ce{VO2}(B) (mp-541404) (\ac{percdistmax}~=~1.75~$\text{\AA}$ with \ac{percdisp}~=~1.13~$\text{\AA}$). From the results obtained from previously studied \ac{RAZIB} cathodes, a \ac{percdisp} limit of 0.7~$\text{\AA}$ was then chosen as the maximum displacement for which the \ac{percdistmax} would be analyzed, as the \ac{percdisp} cap would still allow for realistic percolation paths to be established inside the host structures. The criteria for the evaluation of the \ce{Zn^{2+}} percolation path was then established to be of a \ac{percdistmax} between 1.7~$\text{\AA}$ and 2.5~$\text{\AA}$ for a \ac{percdisp} of less than 0.7~$\text{\AA}$ (\ac{percdistmax}(\ac{percdisp}~$<$~0.7~$\text{\AA}$)). The number of remaining viable candidates was then reduced to 195 materials (see Fig.~\ref{fig:ScreeningFunnel}), with only materials that did not already have \ce{Zn} atoms in the structure being considered during the \ce{Zn^{2+}} percolation path screening process.

The screening criteria for the evaluation of \ac{percdistmax} is also supported by the \ac{Eb} calculation results (see Fig.~S4), where it can be seen that the lower threshold amount of 1.7~$\text{\AA}$ for \ac{percdistmax} is expected to return a \ac{Eb} of approximately 1.5~eV, which is similar to what has been calculated for \ce{V2O5} (mp-25279), a materials previously explored as cathode for \acp{RAZIB}\cite{Fan_PMS_149_2025}. The results also show a trend towards a considerable increase in the calculated \ac{Eb} for materials with \ac{percdistmax} lower than 1.7~$\text{\AA}$, which indicates that further reduction of the \ac{percdistmax} parameter is not appropriate, as materials with overall unfeasible associated energy for \ce{Zn^{2+}} percolation would continue to be considered. On the other hand, considerably low \ac{Eb} values were predicted for structures with higher \ac{percdistmax} associated values (\textit{e.g.}, \ac{Eb}(\ac{percdistmax} = 3.2~$\text{\AA}$) $\approx$ 0.3~eV). The reduced \ac{Eb} indicates a more energetically favourable \ce{Zn^{2+}} percolation within the crystal cell, but the higher distance from the percolation path to the atoms in the crystal structure, as captured by \ac{percdistmax}, indicates that the bond established by \ce{Zn^{2+}} after intercalation would need to be considerably longer. The unavailability of closer atoms for \ce{Zn} to bond to may cause the establishment of an energetically unfavourable coordination environment, which has been shown to promote the degradation of \ac{RAZIB} batteries during cycling\cite{Miliante_EESBat_2_2026}. Therefore, an upper limit for \ac{percdistmax} screening is also necessary in order to avoid the proposal of cathode materials for which the \ce{Zn^{2+}} intercalation would then be expected to promote cathode material degradation, with a maximum value for \ac{percdistmax}(\ac{percdisp}~$<$~0.7~$\text{\AA}$) of 2.5~$\text{\AA}$ being then considered here. It should be noted that the utilization of \ac{percdistmax} is not being proposed here as a direct substitution for the calculation of \ac{Eb} through \ac{NEB} methods, for example, in all future evaluations of ionic percolation in electrode materials, but solely as one more tool to be employed by researchers for expediting the discovery of materials with necessary, but potentially not sufficient, crystal structure geometry for viable ionic percolation.

Finally, the oxidation states of the transition metal centres in the remaining candidates were investigated to determine the materials that were indeed chemically feasible for operation as a cathode. 130 materials were concluded to be viable candidate \ac{RAZIB} cathode materials due to their combined high electrochemical stability, feasibility of \ce{Zn^{2+}} percolation within the structure and practical metallic centre redox couple (see Fig.~\ref{fig:ScreeningFunnel}). From the 130 materials selected for \ac{EZn} calculation, 81 of them were polyanions, and 49 were oxides, with the majority of the materials having \ce{V} (47) or \ce{Mo} (23) as their metal centres (see Fig.~S6). Figure~\ref{fig:ElectStabxPercPath}b shows the distribution of the selected materials for \ac{EZn} calculation with respect to \ac{avgdGpbx} and \ac{percdistmax}(\ac{percdisp}~$<$~0.7~$\text{\AA}$). The calculated \ac{percdistmax}(\ac{percdisp}~$<$~0.7~$\text{\AA}$) for \ce{PWO5} (mp-554866) (\ac{percdistmax}(\ac{percdisp}~$<$~0.7~$\text{\AA}$)~=~2.78~$\text{\AA}$) was higher than the screening limit 2.5~$\text{\AA}$ (see Figure~\ref{fig:ElectStabxPercPath}b). However, \ce{PWO5} was still considered for \ac{EZn} calculation, since it was determined that a percolation path with lower \ac{percdisp} than 0.7~$\text{\AA}$ is able to be established in the structure and also be within the \ac{percdist} screening range (1.7~$\text{\AA}$~$<$~\ac{percdist}~$<$~2.5~$\text{\AA}$). The data compiled for all materials used in the analysis of the electrochemical stability, \ce{Zn^{2+}} ion percolation path, and chemical feasibility for cathode operation is available  \href{https://doi.org/10.5281/zenodo.18566365}{online}, allowing other researchers to implement different screening parameter limits for the exploration of novel \ac{RAZIB} cathodes.

\subsection{Evaluation of the \ce{Zn^{2+}} (de)intercalation potential}

\begin{figure}
\includegraphics{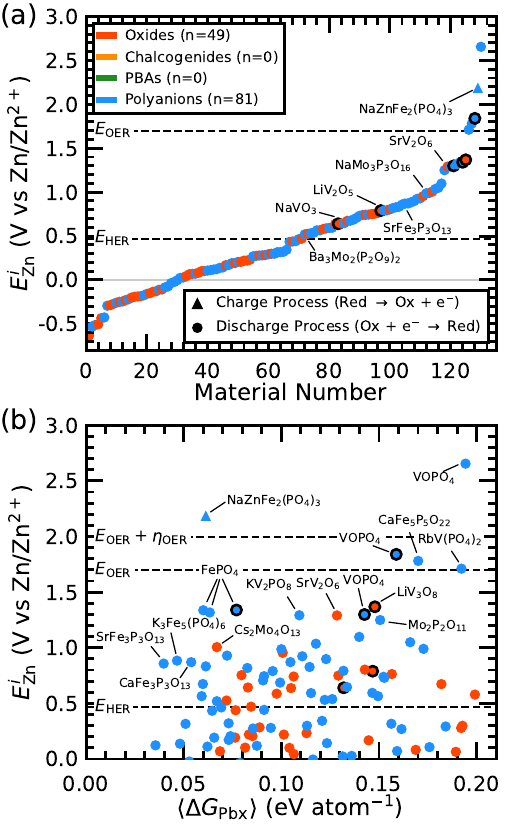}
\caption{Calculated \acf{EZninitial} results for all materials plotted (a) in the order of increasing potential and (b) with respect to the material \acf{avgdGpbx}. Materials that have been previously experimentally investigated are highlighted in the graphs with an outer black line on their markers. The dashed lines represent the reversible potentials for \ac{HER} ($\acs{elect_pot}_{\text{HER}}$) and \ac{OER} ($\acs{elect_pot}_{\text{OER}}$) calculated at a pH~=~5 and [\ce{Zn^{2+}}]~=~1~M, and the practical potential for \ac{OER} considering the overpotential necessary to drive the reaction ($\acs{elect_pot}_{\text{OER}}$ + \acs{overpotential_OER}).}
\label{fig:IntercalationPotential}
\end{figure}

The \ac{EZninitial} calculated for all remaining candidate materials is presented in Figure~\ref{fig:IntercalationPotential}. First, the literature was queried to identify which of the screened materials had been previously experimentally investigated as a cathode for \acp{RAZIB} to validate the \ac{EZninitial} prediction capability. In total, 6 materials (3 oxides and 3 polyanions) were found to have had previous experimental reports, with these materials being highlighted in Figure~\ref{fig:IntercalationPotential} with an outer black line on their markers. The experimental intercalation potential for these materials are presented in Table~\ref{tbl:IntPotentComp}, alongside the theoretical results calculated in this study. Overall, strong agreement can be seen between the experimental and theoretical \ce{Zn^{2+}} intercalation potential results, with absolute errors of less than 0.2~V being seen for all materials. The agreement between the theoretical and experimental results, independently of the class, validates the predictive capability of the methodology employed in this study for the calculation of \ac{EZninitial}. The agreement also allows for a more substantiated analysis of the calculated \ac{EZninitial} results for the previously unexplored results and the proposal of novel materials for experimental investigation as \ac{RAZIB} cathodes.

\setlength\extrarowheight{2pt}
\setlength{\tabcolsep}{1pt}
\begin{table}
  \caption{Experimental and theoretical (this work) initial \ce{Zn^{2+}} intercalation potential results for screened materials that have already been experimentally investigated. Experimental results are approximate and will reflect the expected potential associated with the initial \ce{Zn^{2+}} intercalation.}
  \label{tbl:IntPotentComp}
  \begin{tabular}{ccc}
  \hline
    \multirow{2}{*}{Material} & \multicolumn{2}{c}{\ce{Zn^{2+}} Intercalation Potential (V~vs~\ce{Zn}/\ce{Zn^{2+}})} \\ \cline{2-3}
    ~ & Experimental & Theoretical \\ \hline
    $\alpha$-\ce{VOPO4} (mp-559299) & 1.7 \cite{Verma_AEM_2_2019} & 1.84 \\ 
    \ce{LiV3O8} (mp-27503) & 1.3 \cite{He_ESM_29_2020, Wu_AM_36_2024} & 1.37 \\ 
    \textit{olivine}-\ce{FePO4} (mp-20361) & 1.2 \cite{Shimizu_CEC_11_2024} & 1.34 \\ 
    $\delta$-\ce{VOPO4} (mp-554181) & 1.5 \cite{Zhao_CS_14_2023} & 1.30 \\ 
    \ce{LiV2O5} (mp-19408) & 0.9 \cite{Xiao_CEJ_489_2024} & 0.79 \\ 
    \ce{NaVO3} (mp-555665) & 0.8 \cite{Lu_JAC_975_2024} & 0.64 \\ \hline
\end{tabular}
\end{table}

For the \ce{Zn^{2+}} (de)intercalation in a cathode material to be energetically favourable, it is necessary for \acf{HZn} to be negative, which would then lead to positive \ac{EZninitial} values (see Eqs.~\eqref{eqn:HZn}~and~\eqref{eqn:EZn}). However, 30 of the investigated materials reported positive \ac{HZn} values, indicating that \ce{Zn^{2+}} would not intercalate into the materials at practical potentials for operation as cathodes for \ac{RAZIB}. These materials can be identified in Figure~\ref{fig:IntercalationPotential}a by having their reported potentials be less than 0~V~vs~\ce{Zn}/\ce{Zn^{2+}}. Also, 40 other materials reported positive \ac{EZninitial} that were less than the reversible potential for \ac{HER}, deeming them to be impractical for operation in cathodes for \acp{RAZIB} due to the low associated potential for \ce{Zn^{2+}} intercalation and the risk of electrolyte decomposition via the \ac{HER}. The \ac{OER}, on the other hand, has a higher associated reversible potential and requires greater \ac{overpotential} for driving the electrochemical reaction than \ac{HER}\cite{Wang_NRE_2_2023, Yu_Small_19_2023, Zhang_JECE_2025, Arshad_SusMat_5_2025, Yan_IJHE_228_2026}. For example, leading catalysts for \ac{OER} require an \ac{overpotential} higher than 250~mV to obtain a current density of approximately 10~mA~cm$^{-2}$\cite{Wang_NRE_2_2023, Yu_Small_19_2023, Zhang_JECE_2025}. Therefore, it is expected that materials which reported a \ac{EZninitial} lower than the reversible potential for \ac{OER} ($\acs{elect_pot}_{\text{OER}}$) plus an associated \ac{overpotential_OER} of 300~mV (\textit{i.e.}, \ac{EZninitial} $<$ $\acs{elect_pot}_{\text{OER}}$ + \ac{overpotential_OER}) would still be able to operate as cathode materials for \acp{RAZIB} without potential risk of electrolyte degradation. Only 2 materials were then considered not practical for utilization as active cathode materials due to the possibility of oxygen evolution during \ac{RAZIB} cycling (\textit{i.e.}, \ac{EZninitial} $\geq$ $\acs{elect_pot}_{\text{OER}}$ + \ac{overpotential_OER}). To understand why more than half of the materials investigated reported \ac{EZninitial} as impractical for application as a \ac{RAZIB} cathode, the oxidation state for the metal centres and the chemical environment for \ce{Zn^{2+}} (de)intercalation were investigated.

It has long been established in the \ac{LIB} literature the direct relationship between higher metallic centre oxidation state and higher battery operation potentials for redox couples with the same crystal structure (\textit{i.e.}, \ac{elect_pot}(\textit{Mtl}$^{\text{IV}}_{\text{(s)}}$/\textit{Mtl}$^{\text{III}}_{\text{(s)}}$) will be greater than  \ac{elect_pot}(\textit{Mtl}$^{\text{III}}_{\text{(s)}}$/\textit{Mtl}$^{\text{II}}_{\text{(s)}}$) for the same \textit{Mtl} species)\cite{Goodenough_CM_22_2010, Padhi_UT_1997}. The direct relationship between metallic oxidation state and intercalation potential was also demonstrated by our group in the investigation of binary oxides as cathodes for \acp{RAZIB}\cite{Miliante_JPCC_128_2024}. When evaluating the \ac{EZninitial} results for the polyanions and oxide materials investigated here, a similar trend between oxidation state and \ce{Zn^{2+}} (de)intercalation potential was observed, as shown in Figure~S7. The trend can be clearly seen for metals for which the \ac{EZninitial} was calculated for materials at varied oxidation states, such as \ce{V} and \ce{Mo}. The results show that materials with high average oxidation states (\textit{e.g.}, \ce{V^{5+}_{(s)}} in \ce{LiV3O8} (mp-27503) and  \ce{KV2PO8} (mp-557947), and \ce{Mo^{5.33+}_{(s)}} in \ce{LiMo3P3O16} (mp-17314)) reported potentials greater than 1~V~vs~\ce{Zn}/\ce{Zn^{2+}} while materials with lower average oxidation states (\textit{e.g.}, \ce{V^{3+}_{(s)}} in \ce{KVP2O7} (mp-16812), and \ce{Mo^{3+}_{(s)}} in \ce{LiMoP2O7} (mp-18987)) tend to return lower and even negative \ac{EZninitial} values (see Fig.~S7). Also, metallic redox couples previously identified as being impractical for \ac{RAZIB} cathode application due to their low \ce{Zn^{2+}} intercalation potential\cite{Miliante_JPCC_128_2024} (\textit{e.g.}, metal oxidation state at \ac{mchg} equal to \ce{Ti^{4+}_{(s)}}, \ce{V^{3+}_{(s)}}, \ce{Cr^{3+}_{(s)}}, \ce{Mo^{3+}_{(s)}}, or \ce{W^{5+}_{(s)}}) were also identified here as having negligible \ac{EZninitial}.

However, the calculated \ac{EZninitial} results can not be explained solely from the metal centres oxidation states, as some materials do not follow the expected trend. For example, \ce{RbVPHO6} (mp-1201601) (\ce{V^{5+}_{(s)}}, 0.12~V~vs~\ce{Zn}/\ce{Zn^{2+}}), \ce{Rb2MoO4} (mp-19212) (\ce{Mo^{6+}_{(s)}}, -0.29~V~vs~\ce{Zn}/\ce{Zn^{2+}}), and \ce{Li3VO4} (mp-19219) (\ce{V^{5+}_{(s)}}, 0.06~V~vs~\ce{Zn}/\ce{Zn^{2+}}) have high average metallic oxidation state, but return very low \ac{EZninitial}. In order to explain why these and other materials do not follow the expected metallic oxidation state trend, and also why the calculated \ac{EZninitial} results are scattered through a wide potential range (see Fig.~\ref{fig:IntercalationPotential}), it is necessary to look into the chemical environments established after \ce{Zn^{2+}} intercalation. \ce{Zn} atoms are preferably bonded in inorganic structures through a tetrahedral or octahedral coordination, as a result of the complete filling of the degenerate molecular orbitals by the \ce{Zn^{2+}} $d^{10}$ electrons \cite{Barak_Zinc_1993, Neumann_MMJMS_28_1949, Wedepohl_HandGeo_1_1969, Jean_MOTMC_2005}. When investigating the \ce{Zn^{2+}} bonding in the materials which did not follow the oxidation state trend, it is possible to attest that the coordination environments established after the \ac{MD} relaxation are not energetically favourable. For example, the intercalated \ce{Zn^{2+}} ion is bonded to  \ce{RbVPHO6} through a planar coordination (Fig.~S8a,b), while angular and distorted tetrahedral coordinations were found for \ce{Rb2MoO4} (Fig.~S8c,d) and \ce{Li3VO4} (Fig.~S8e,f), respectively. On the other hand, structures which reported high \ac{EZninitial} and followed the expected trend with respect to the metallic oxidation state had the intercalated \ce{Zn^{2+}} atom in an energetically favourable coordination environment. For example, the energetically favourable tetrahedral coordination was established in the relaxed structures of \ce{KV2PO8} (mp-557947) (\ce{V^{5+}_{(s)}}, 1.30~V~vs~\ce{Zn}/\ce{Zn^{2+}}) (Fig.~S9a,b), \ce{Cs2Mo4O13} (mp-1202511) (\ce{Mo^{6+}_{(s)}}, 1.00~V~vs~\ce{Zn}/\ce{Zn^{2+}}) (Fig.~S9c,d), \ce{BaV2O6} (mp-18929) (\ce{V^{5+}_{(s)}}, 0.95~V~vs~\ce{Zn}/\ce{Zn^{2+}}) (Fig.~S9e,f), demonstrating a direct relationship between high \ac{EZninitial} and favourable \ce{Zn^{2+}} coordination environment in the structure. However, the consideration of a suitable \ce{Zn^{2+}} coordination environment is often overlooked during the investigation of novel cathode materials, with researchers proposing materials with considerably large interlayer spacing and tunnels which will not necessarily support stable \ce{Zn^{2+}} intercalation\cite{Tang_JMCA_7_2019, Dai_AEM_2025}. From the analysis of the \ac{EZninitial} results for the investigated materials, it is possible to conclude that to realize \acp{RAZIB} with high \ce{Zn^{2+}} intercalation potential, it is important to screen for novel cathode materials that have redox active species at high oxidation states, and also a crystal structure that allows for the establishment of energetically favourable \ce{Zn^{2+}} coordinations after intercalation. It is important to note that the considerations presented here (\textit{i.e.}, high oxidation state for the redox active species, and favourable structural environment for forming stable ion intercalation) are also applicable for the discovery of novel cathode materials with high operating potentials in other intercalating ion battery chemistries, since the conclusions obtained from the \ac{EZn} analysis are independent of the investigated \ac{RAZIB} chemistry. For other intercalating ions (\textit{e.g.}, \ce{Na^{+}}, \ce{Ca^{2+}}), the energetically favourable ion coordination environment will not necessarily be the tetrahedral and octahedral coordination environments established for \ce{Zn^{2+}}.

In total, 58 materials were found to have suitable electrochemical stability, structural availability for \ce{Zn^{2+}} percolation, chemically reasonable transition metal oxidation state and \acl{EZninitial} (\textit{i.e.}, $\acs{elect_pot}_{\text{HER}}$ $<$ \ac{EZn} $<$ $\acs{elect_pot}_{\text{OER}}$ + \acs{overpotential_OER}) for use as a cathode in \acp{RAZIB}. As presented in Table~\ref{tbl:IntPotentComp}, 6 out of the 58 materials identified have already been previously explored as \ac{RAZIB} cathodes, but no reports were found for the remaining materials. The majority of the newly identified materials (45 out of 52) report considerably low \ac{EZninitial} (lower than~1.2~V~vs~\ce{Zn}/\ce{Zn^{2+}}), deeming them of low scientific interest as \ac{RAZIB} cathodes. However, the remaining 7 materials display significant \ac{EZninitial} and high electrochemical stability, which warrants a deeper investigation as potential next-generation \ac{RAZIB} cathode materials. The 7 shortlisted materials from the screening methodology are presented in Table~S1, alongside their respective calculated \acf{EZninitial}, \acf{EZnavg}, \acf{avgdGpbx}, \acf{Qw}, and \acf{Ww}. 

Two polymorphs of \ce{FePO4} were identified through the theoretical screening ($\alpha$-\ce{FePO4} (mp-19109) and $\beta$-\ce{FePO4} (mp-19752)) with elevated predicted \acl{EZninitial} (\ac{EZninitial}~$>$~1.3~V~vs~\ce{Zn}/\ce{Zn^{2+}}) and high electrochemical stability (\acs{avgdGpbx}~=~0.06~eV atom$^{-1}$), alongside high \acl{Qw} ($>$~170~mA~h~g$^{-1}$). Both $\alpha$- and $\beta$-\ce{FePO4} are also of further interest for experimental investigation due to the previous implementation of both \textit{olivine}-\ce{FePO4} and amorphous \ce{FePO4} as active materials in cathodes for \acp{RAZIB}\cite{Mathew_NPGAM_6_2014, Shimizu_CEC_11_2024}. Therefore, to probe the experimental viability of these new \ce{FePO4} polymorphs as cathode materials for \acp{RAZIB}, $\alpha$-\ce{FePO4} was experimentally synthesized and submitted to electrochemical testing in battery cells.

The \ac{XRD} diffractogram for the $\alpha$-\ce{FePO4} sample is shown in Figure~S10, where the successful synthesis of the $\alpha$-\ce{FePO4} phase can be verified, alongside impurities from the precursors utilized during the synthesis process (\ce{Fe2O3} and \ce{(NH4)2HPO4}). A \ac{RAZIB} coin cell employing $\alpha$-\ce{FePO4} as the active cathode material was then assembled, utilizing a 1~M \ce{ZnSO4} solution as the electrolyte and zinc metal as the anode (detailed experimental setup is presented in the \ac{SI}). First, the electrochemical response of the $\alpha$-\ce{FePO4} sample as a cathode material for \acp{RAZIB} was investigated through \ac{CV} experiments. Figure~S11 presents the results for the \ac{CV} tests, where only one discernible oxidative peak close to 1.2~V~vs~\ce{Zn}/\ce{Zn^{2+}} was able to be identified, being present at all scan rates tested. The \ac{CV} obtained here for $\alpha$-\ce{FePO4} matches the  previously captured \ac{CV} for amorphous \ce{FePO4} cathode materials for \acp{RAZIB} in 1~M \ce{ZnSO4} electrolyte \cite{Mathew_NPGAM_6_2014}, as a single main peak was also observed around 1.2~V~vs~\ce{Zn}/\ce{Zn^{2+}} during the anodic sweep. However, the current density for the oxidative peak obtained from the \ac{CV} experiments of $\alpha$-\ce{FePO4} is orders of magnitude lower than for amorphous \ce{FePO4}\cite{Mathew_NPGAM_6_2014}, indicating that the assembled $\alpha$-\ce{FePO4} cathode has a significantly lower Faradaic response in the potential window and time scales tested. To evaluate the performance of the $\alpha$-\ce{FePO4} cathode during battery cycling, \ac{GCD} experiments were conducted, with the results being shown in Figure~S12. Overall, no charge storage plateaus are able to be identified from the \ac{GCD} results for the synthesized $\alpha$-\ce{FePO4} sample as active cathode material, indicating the inability of the assembled battery cell to cycle. 

The pursued electrochemical tests for the $\alpha$-\ce{FePO4} sample paint a puzzling picture with respect to the use of this phase as a cathode material for \acp{RAZIB}. On one hand, the \ac{CV} experiment results for $\alpha$-\ce{FePO4} display the presence of the anodic peak at 1.2~V~vs~\ce{Zn}/\ce{Zn^{2+}} (Fig.~S11), matching the \ac{CV} results previously reported for amorphous \ce{FePO4}\cite{Mathew_NPGAM_6_2014} and possibly pointing to a similar redox mechanism for both active materials. However, when tested with respect to the battery cycling through \ac{GCD} experiments (Fig.~S12), the assembled $\alpha$-\ce{FePO4} cathode did not display the ability to meaningfully store charge, as no plateaus were observed in the \ac{GCD} cycling profile, differing from the results for amorphous \ce{FePO4}\cite{Mathew_NPGAM_6_2014}.

\begin{figure}
\includegraphics{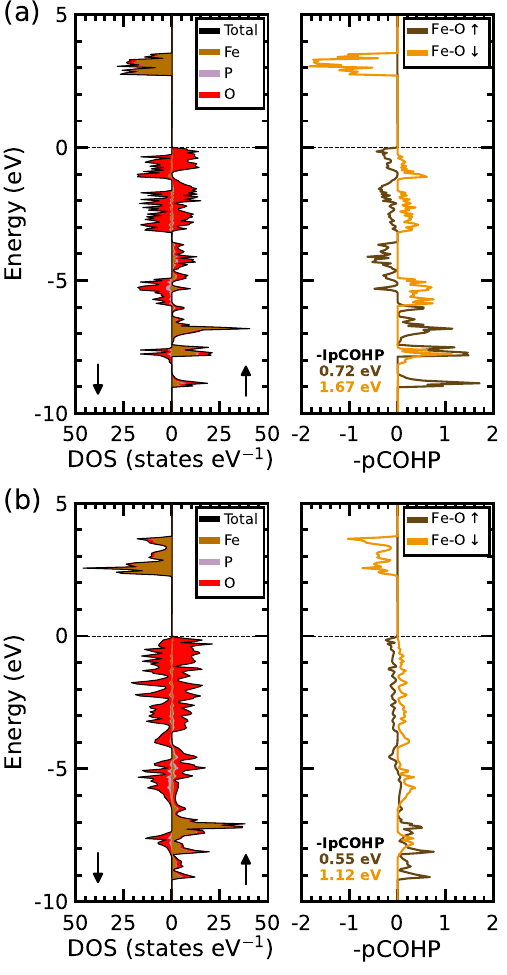}
\caption{\Acf{DOS} and \acf{PDOS} results for (a) $\alpha$-\ce{FePO4} (mp-19109) and (b) \textit{olivine}-\ce{FePO4} (mp-20361) with the associated \acf{pCOHP} analysis results for the \ce{Fe}-\ce{O} bond in the structure. Energy is shifted so the \acf{Ef} is at 0~eV. Insets give the average \acf{-IpCOHP} per \ce{Fe}-\ce{O} bond.}
\label{fig:FePO4_COHP}
\end{figure}

To explore possible differences between $\alpha$-\ce{FePO4} and \textit{olivine}-\ce{FePO4}, the latter having been reported as an active \ac{RAZIB} cathode\cite{Shimizu_CEC_11_2024}, we compared their \ce{Fe}-\ce{O} bonding through \ac{pCOHP} calculations. The calculated \ac{DOS}, \ac{PDOS} and \ac{pCOHP} results for each material are presented in Figure~\ref{fig:FePO4_COHP}. The \ac{PDOS} of the two polymorphs are similar, with the valence states close to the \ac{Ef} having a strong \ce{O} character and the lower conduction band states being mainly composed of \ce{Fe} orbitals. The \ac{-IpCOHP} integrated up to the \ac{Ef} gives the contribution of each \ce{Fe}-\ce{O} pair to the band energy. In $\alpha$-\ce{FePO4}, the four \ce{Fe}-\ce{O} bonds of the \ce{FeO4} tetrahedron are nearly equivalent (1.87 to 1.88~\AA), with an average \ac{-IpCOHP} of 0.72 and 1.67~eV per bond for the spin-up and spin-down channels, respectively. In \textit{olivine}-\ce{FePO4}, the six bonds of the \ce{FeO6} octahedron are longer (1.93 to 2.15~\AA) and considerably more heterogeneous, with \ac{-IpCOHP} values from 0.36 to 0.75~eV (spin up) and from 0.78 to 1.48~eV (spin down), and averages of 0.55 and 1.12~eV per bond. The larger \ac{-IpCOHP} per bond in $\alpha$-\ce{FePO4} is consistent with its shorter bonds and lower \ce{Fe} coordination number. When summed over the coordination shell of each \ce{Fe} atom, however, the values of the two polymorphs are nearly identical (2.9 and 6.7~eV for $\alpha$-\ce{FePO4}, and 3.3 and 6.7~eV for \textit{olivine}-\ce{FePO4}). The two polymorphs therefore differ in how the \ce{Fe}-\ce{O} bonding is distributed over the coordination shell rather than in its total magnitude, and the present calculations do not establish a link between these differences and the electrochemical inactivity of $\alpha$-\ce{FePO4}.

The same analysis was performed for $\beta$-\ce{FePO4}, the other \ce{FePO4} polymorph identified in the screening (Fig.~S13). Similar to $\alpha$-\ce{FePO4}, it contains \ce{FeO4} tetrahedra, and its \ac{-IpCOHP} values are similar to those of $\alpha$-\ce{FePO4} in both spin channels. The comparison of \ac{-IpCOHP} values is meaningful here only because the three systems share the same composition, and it is not possible to derive a general \ac{-IpCOHP} criterion for \textit{Mtl}-\ce{O} bonds from these results.

The behaviour of $\alpha$-\ce{FePO4} in lithium- and sodium-ion batteries is consistent with a limited ability of this structure to accommodate intercalating ions\cite{Okada_JPS_146_2005, Song_MRB_37_2002, Shi_ESSL_8_2005, Wang_PCCP_17_2015}. For example, \citet{Okada_JPS_146_2005} evaluated the performance of amorphous \ce{FePO4} and crystalline $\alpha$-\ce{FePO4} as cathodes for \acp{LIB}, demonstrating that the \ce{Li+} ions which intercalated into $\alpha$-\ce{FePO4} during the discharge cycles did not fully deintercalate during subsequent charge. The battery was observed to lose capacity during cycling due to the irreversibility of the \ce{Li^{+}} (de)intercalation process, with \ce{Fe^{II}_{(s)}} sites with reduced oxidation state still identified to be present in the cathode material even after full battery charge through $^{57}\ce{Fe}$ Mössbauer spectroscopy\cite{Okada_JPS_146_2005}. The monovalent \ce{Li^{+}} ion is then able to intercalate within the $\alpha$-\ce{FePO4} crystal to a considerable extent during discharge, but the (de)intercalation process is not able to proceed reversibly.

It is worth noting that the screening methodology and the bonding analysis presented in this work are based on bulk properties. The electrode-electrolyte interface will also affect the electrochemical response of the cathode during battery cycling, with the dynamics for \ce{Zn^{2+}} transfer from the electrolyte to the electrode crystal structure and the potential passivation of the electrode surface being directly influenced by the established interface. While of high relevance for the investigation of the performance of the cathode material during battery operation, the atomistic modelling of the electrode-electrolyte interface is computationally very challenging and expensive, due to the complexities of capturing the interaction between two different materials in distinct phases of matter. For example, \textit{ab initio} simulations of battery materials interfaces are commonly run for only short periods of time (\textit{i.e.}, picoseconds long), allowing the researcher to investigate, for example, the structural stability of the proposed electrode materials at the expected conditions of battery operation\cite{Nirjhar_CSA_684_2024, Fardi_AMI_11_2024}. However, longer-lasting dynamics, such as ionic intercalation, phase segregation, or the formation of a surface passivating phase, are not currently able to be accurately captured due to the associated computational costs of running such atomistic simulations. Considerable research interest is currently being devoted to establishing methods to expand machine learning interatomic potentials to also describe solid-liquid interfaces under applied external potential, which would considerably speed up the evaluation of the electrode-electrolyte interface in battery materials. However, current leading methodologies have only been shown to describe simpler interfaces (\textit{e.g.}, \ce{Pt}(111)-\ce{H2O}\cite{Bergmann_JCP_164_2026, Chen_JCTC_22_2026}, \ce{Cu}(100)-\ce{H2O}\cite{Bergmann_PRL_135_2025}), with considerable research still necessary to allow for the accurate simulation of the interaction in more complex systems, for example between $\alpha$-\ce{FePO4} and a 1~M \ce{ZnSO4} aqueous solution, which is of interest to this study. The present computational cost and complexity for the modelling of battery interfaces also explain why their consideration has been overlooked in previous articles focused on searching for novel cathode materials for \acp{RAZIB}\cite{Cai_ESM_42_2021, Wudil_AAMI_17_2025, Zhou_ATS_4_2021}. Therefore, the consideration of interfacial effects during the atomistic screening and research of battery cathode materials is still an open research opportunity, being of considerable scientific relevance due to the potential impact of the electrode-electrolyte interface on experimental electrochemical performance.

The electrochemical stability criterion used in this work favours hosts that are insoluble in the aqueous electrolyte. As a consequence, the workflow filters out materials whose charge storage relies on dissolution and redeposition of the active material, a mechanism reported for \ce{MnO2} cathodes \cite{Moon_AS_8_2021, Li_ACE_136_2024, Sanders_CM_38_2026, Shang_NC_17_2026}. The calculated Pourbaix diagram in Fig.~\ref{fig:MethodExample}c shows that iron has no stable aqueous species within the operating window, which excludes this energy storage mechanism for \ce{FePO4}.

\subsection{Proposed materials for experimental testing}

 The final list of materials proposed here for experimental testing is presented in Table~\ref{tbl:ProposedMaterials}, and it contains three materials, \ce{CaFe5P5O22}, \ce{RbV(PO4)2}, and \ce{Mo2P2O11}. From the initial shortlist presented in Table~S1, the \ce{FePO4} polymorphs were discarded due to the experimental failure of $\alpha$-\ce{FePO4}. Another two materials in \ce{KV2PO8} (mp-557947) and \ce{SrV2O6} (mp-19038) were disregarded from endorsement for experimental testing due to the considerable difference in \ce{Zn^{2+}} intercalation potential between \ac{EZninitial} and \ac{EZnavg}, indicating that \ac{EZninitial} would not be sustained through the discharge cycle and a limited potential could be obtained from the cathode materials during operation. On the other hand, the materials being proposed here (Tab.~\ref{tbl:ProposedMaterials}) have notably higher \ac{EZnavg} and \ac{Ww}, granting them to be more practical  as novel cathode materials for \acp{RAZIB}. Moreover, in \ce{CaFe5P5O22}, the only \ce{Fe}-containing phosphate proposed here, the \ce{Fe} atoms are octahedrally coordinated to \ce{O} (\ce{FeO6}), as in the electrochemically active \textit{olivine}-\ce{FePO4}, rather than tetrahedrally coordinated as in the inactive $\alpha$-\ce{FePO4}.

\vspace*{\fill}
\setlength\extrarowheight{2pt}
\setlength{\tabcolsep}{3pt}
\begin{table}
  \caption{Selected materials proposed for experimental testing, alongside their calculated \acf{EZninitial}, \acf{EZnavg}, \acf{avgdGpbx}, \acf{Qw}, and \acf{Ww}.}
  \label{tbl:ProposedMaterials}
  \begin{tabular}{c|cccc}
  \hline
  \multirow{2}{*}{Material}  &~~~~\acs{EZninitial}~~~~~\acs{EZnavg}~~~& \acs{avgdGpbx} & \acs{Qw} & \acs{Ww} \\ 
 &  (V~vs~\ce{Zn}/\ce{Zn^{2+}})  & (eV~atom$^{-1}$) & (mA~h~g$^{-1}$) & (W~h~kg$^{-1}$) \\ \hline
\ce{CaFe5P5O22} (mp-818432)	& ~~~~1.78 ~~~~~ 1.30 ~~~ & 0.17 & 162.2 & 210.9 \\
\ce{RbV(PO4)2} (mp-818600)	& ~~~~1.71 ~~~~~ 1.50 ~~~ & 0.19 & 82.1 & 123.2 \\
\ce{Mo2P2O11} (mp-636950) & ~~~~1.25 ~~~~~ 0.96 ~~~ & 0.15 & 124.7 & 119.9 \\
\hline
\end{tabular}
\end{table}
\vspace*{\fill}

The materials proposed in this work represent a meaningful contribution to expanding the number of viable cathode materials for \acp{RAZIB}, providing experimentally synthesizable candidates with promising electrochemical stability, \ce{Zn^{2+}} intercalation potential, theoretical gravimetric capacity, and energy density as identified here through atomistic-level theoretical screening. However, it is important to emphasize that the metrics evaluated in this study are necessary, but not sufficient conditions for the successful experimental realization of the proposed materials as \ac{RAZIB} cathodes for grid-scale applications. Parameters such as rate capability, cycling stability at low C-rates, high mass loading, and scalability of the manufacturing process are inherently dependent on experimental synthesis conditions, electrode formulation, and cell design, and cannot be probed at the size and time scales accessible through the atomistic simulations pursued here. Experimental optimization of the proposed materials is then essential to realize their ultimate potential as active cathode materials in \acp{RAZIB} for grid-scale energy storage applications. Therefore, the present work advances \ac{RAZIB} development by identifying promising next-generation cathode materials, while acknowledging that experimental challenges remain to be overcome before the commercialization of this technology can be realized.

\section{Conclusion}

\acresetall

In this study, a comprehensive screening of previously synthesized oxides, chalcogenides, \acp{PBA} and polyanions as novel \ac{RAZIB} cathodes was performed. The data from more than 2000 experimentally confirmed materials were retrieved from the Materials Project database, with their structural availability for \ce{Zn^{2+}} percolation, electrochemical stability in aqueous media and chemical feasibility for use as a cathode being evaluated. In total, 130 materials were identified through the screening process as having suitable structural, electrochemical and chemical characteristics for operation as a \ac{RAZIB} cathode, for which the \ac{EZninitial} was then calculated. Higher \ac{EZninitial} results were revealed to be obtained for materials that had transition metals at higher oxidation states in their structure prior to \ce{Zn^{2+}} intercalation and were also able to accommodate the intercalating \ce{Zn^{2+}} ion in a favourable coordination environment. The determination of a relationship between metal oxidation state and ion coordination to \ac{EZn} presented in this study contributes to future exploration of cathode materials with high intercalation \ce{Zn^{2+}} potentials for \acp{RAZIB}, as well as other battery chemistries.

A negligible Faradaic response was observed in \ac{CV} and \ac{GCD} experiments for $\alpha$-\ce{FePO4}, one of the initially shortlisted materials, which identifies it as a false positive of the screening protocol. A comparison with \textit{olivine}-\ce{FePO4} through \ac{pCOHP} calculations showed that the total \ce{Fe}-\ce{O} bonding per \ce{Fe} atom is nearly the same in both polymorphs, while its distribution over the coordination shell differs. None of the descriptors considered here was able to identify $\alpha$-\ce{FePO4} as an inactive host, and a screening criterion capable of isolating such false positives remains to be established. Finally, three previously unexplored materials (\ce{CaFe5P5O22}, \ce{RbV(PO4)2}, and \ce{Mo2P2O11}) with leading metrics for \ac{RAZIB} operation, such as high intercalation potential, electrochemical stability, theoretical gravimetric capacity, and energy density, were here proposed for experimental investigation as \ac{RAZIB} cathodes for the first time, directly supporting the continued development of this battery technology. In addition, the methodology employed in this study for the discovery of novel \ac{RAZIB} cathode materials can also be employed for the exploration of cathode materials for other battery chemistries, especially with aqueous electrolytes, positively contributing to the continuous implementation of rechargeable batteries for the energy transition.

\section{Author Contributions}

\begin{itemize}
\item \textbf{Caio Miranda Miliante}: Conceptualization, Methodology, Software, Validation, Formal analysis, Investigation (Computational), Data curation, Writing - Original Draft, Writing - Review \& Editing, Visualization, and Project administration.
\item \textbf{Yuzhen Deng}: Investigation (Experimental), Writing - Review \& Editing.
\item \textbf{Brian D. Adams}: Writing - Review \& Editing, and Funding acquisition.
\item \textbf{Drew Higgins}: Resources, Writing - Review \& Editing, Supervision, and Funding acquisition.
\item \textbf{Oleg Rubel}: Conceptualization, Resources, Writing - Review \& Editing, Supervision, Funding acquisition.
\end{itemize}

\begin{acknowledgement}

The authors gratefully acknowledge the financial support from  \href{https://salientenergyinc.com/}{Salient Energy Inc.} and the \href{https://www.nserc-crsng.gc.ca/index_eng.asp}{Natural Sciences and Engineering Research Council of Canada (NSERC)} Alliance Program. C.M.M. acknowledges the \href{https://www.alliancecan.ca/en}{Digital Research Alliance of Canada} and \href{https://www.computeontario.ca/}{Compute Ontario} for the computing resources utilized in this research. C.M.M thanks Prof. Xavier Rocquefelte (University of Rennes) for the discussion about the utilization of \acf{COHP} for the analysis of the proposed materials. \Acf{XRD} experiments were conducted in the \href{https://research.science.mcmaster.ca/rcis-and-cores/services/mcmaster-analytical-x-ray-diffraction-facility-max/}{McMaster Analytical X-Ray Diffraction Facility (MAX)}.

\end{acknowledgement}

\begin{suppinfo}

\acresetall

The \ac{SI} for the paper includes figures for: the distribution of queried materials with respect to the class and transition metal centre; \ac{percdistmax} and \ac{avgdGpbx} results for all screened materials separated by material class; \ac{percdistmax} and \ac{avgdGpbx} results for chalcogenide materials colour-coded by chalcogenide centre (\ce{S}, \ce{Se}, and \ce{Te}); results for the \ac{Eb} calculation and their relationship to the calculated \ac{percdistmax}; percolation path associated with \ac{percdistmax} for \ce{Li3VO4} and \ce{Sr2V2O7}; distribution of material selected for \ac{EZn} calculation with respect to their class (oxide or polyanion) and transition metal centre; \ac{EZn} results separated by transition metal centre and colour-coded by average metal centre oxidation state in the material; \ce{Zn^{2+}} coordination environment in \ce{RbVPHO6}, \ce{Rb2MoO4}, \ce{Li3VO4}, \ce{KV2PO8}, \ce{Cs2Mo4O13}, and \ce{BaV2O6} after relaxation; table containing the \ac{EZninitial}, \ac{EZnavg}, \ac{avgdGpbx}, \ac{Qw}, and \ac{Ww} results for the shortlisted materials; \ac{XRD}, \ac{CV}, and \ac{GCD} results for synthesized $\alpha$-\ce{FePO4} material; \ac{DOS}, \ac{PDOS} and \ac{pCOHP} results for $\beta$-\ce{FePO4}, alongside the following citations\cite{Henkelman_JCP_113_2000, Henkelman_JCP_113_2000_b, Jain_AM_1_2013,  Ong_CMS_97_2015, Bull_MA_2_2021, Jain_PRB_84_2011, Dudarev_PRB_57_1998}

The Python code developed for this study is available \href{https://github.com/CMMiliante/ABCs}{online}, with the compiled data for all materials utilized for parameter screening available at a Zenodo \href{https://doi.org/10.5281/zenodo.18566365}{repository}.

\end{suppinfo}

\bibliography{bibliography}

\end{document}



\clearpage

\textbf{\LARGE{Computational Methods}}

\large{\textbf{Calculation of the \acf{Eb}}}

The \acf{Eb} was obtained through \ac{CINEB} calculations of the \ce{Zn^{2+}} percolation through the calculated percolation path for each material. The \ac{CINEB} implementation for \ac{VASP} introduced by \citet{Henkelman_JCP_113_2000}\cite{Henkelman_JCP_113_2000_b} was utilized, with further information being found \href{https://henkelmangroup.github.io/vtsttools}{online}. The theoretical \ac{DFT} background presented in the manuscript was also used for the \ac{Eb} calculations, with any relevant differences being explicitly mentioned here. The number of valence electrons considered for \ce{H}, \ce{Na}, \ce{K}, \ce{Mn}, \ce{V}, \ce{P}, \ce{O}, \ce{Te}, and \ce{Zn} were respectively 1, 7, 7, 13, 11, 5, 6, 6, and 12. Hubbard \textit{U} corrections of 3.9 and 3.1~ev were respective applied to \textit{d} electrons of \ce{Mn} and \ce{V} atoms\cite{Jain_PRB_84_2011, Dudarev_PRB_57_1998}. The energy cutoff was set to 276.7~eV for \ce{K6MnTe4} (mp-18246), and 400~eV for all other materials, matching the maximum energy cutoff from the pseudopotentials being considered in each calculation. The Brillouin zone was sampled through a uniform k-mesh with  20 subdivisions per $\text{\AA}^{-1}$. The unit cell structures for each material, as retrieved from the Materials Project database\cite{Jain_AM_1_2013, Ong_CMS_97_2015}, was considered for the calculation of \ac{Eb}, with full cell and atomic position relaxation being conducted for each structure without and with \ce{Zn^{2+}} intercalated prior to the \ac{CINEB} calculations. The initial \ce{Zn^{2+}} intercalated structures were created by placing a \ce{Zn} atom on the percolation path associated with the \ac{percdistmax}.

\textbf{\LARGE{Experimental Methods}}

\large{\textbf{Materials Synthesis and Characterization}}

An $\alpha$-\ce{FePO4} (mp-19109) sample material was prepared by a solid-state reaction method\cite{Bull_MA_2_2021}. 600~mg of \ce{Fe2O3} ($\geq$ 98$\%$, Sigma-Aldrich) and 1012.6~mg \ce{(NH4)2HPO4} ($\geq$ 99.995$\%$, Sigma-Aldrich) were thoroughly mixed by grinding with a mortar and pestle. Then, the mixed powder was transferred to a combustion boat and heated at 900~$\degree$C for 24 hours in a tube furnace with a heating rate of 5 $\degree$C min$^{-1}$. The crystal structure of the synthesized $\alpha$-\ce{FePO4} was analyzed by \ac{XRD} (Bruker D8 Venture) with a \ce{Cu} K-$\alpha$ (1.5406~$\text{\AA}$) radiation source at 40~kV. 

\large{\textbf{Electrochemical Measurements}}

The slurry of the cathode consisted of $\alpha$-\ce{FePO4} (80 wt.$\%$), acetylene black (10 wt.$\%$), and \ac{PVDF} (10 wt.$\%$), with \ac{PVDF} dissolved in \ac{NMP} solvent. The slurry was cast on carbon paper (AvCarb P50) with a doctor blade and dried in a vacuum oven at 60~$\degree$C for 12 hours. Circular electrodes were then cut from the dried film, with an active material mass loading of around 1.5~mg~cm$^{-2}$. To prepare the full battery for testing, a $\alpha$-\ce{FePO4} cathode (12~mm in diameter), a zinc anode (14.3~mm in diameter), a glass microfiber membrane (16~mm in diameter, VWR), and 1~M \ce{ZnSO4} aqueous electrolyte were assembled in a CR2032 coin-type cell. The \ac{CV} tests were run through a BioLogic VSP-300 Potentiostat workstation at scan rates from 1 to 10~mV~s$^{-1}$, while \ac{GCD} experiments were performed with the use of the NEWARE BTS4000-5V100mA battery testing system at a current density of 50 mA~$g^{-1}$ and a voltage range from 0.5 to 1.5~V~vs~Zn/Zn$^{2+}$.

\clearpage

\vspace*{\fill}
\begin{figure}
\includegraphics[width=0.8\textwidth]{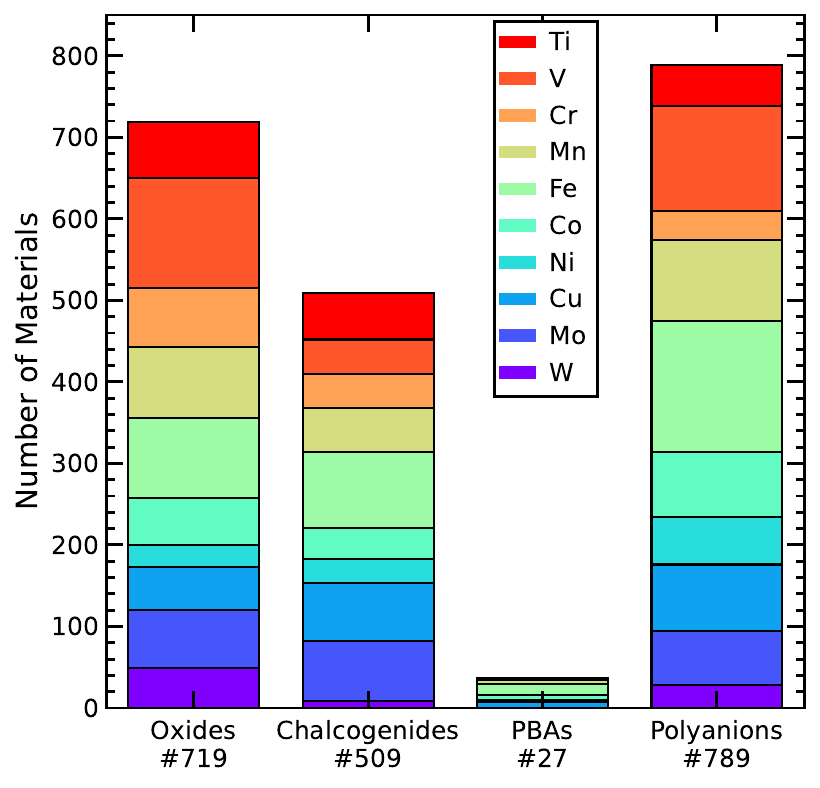}
\caption{Distribution of queried materials from the database for each investigated material class colour-coded by transition metal centre.}
\label{fig:QueriedMaterials}
\end{figure}
\vspace*{\fill}

\vspace*{\fill}
\begin{figure}
\includegraphics[width=1\textwidth]{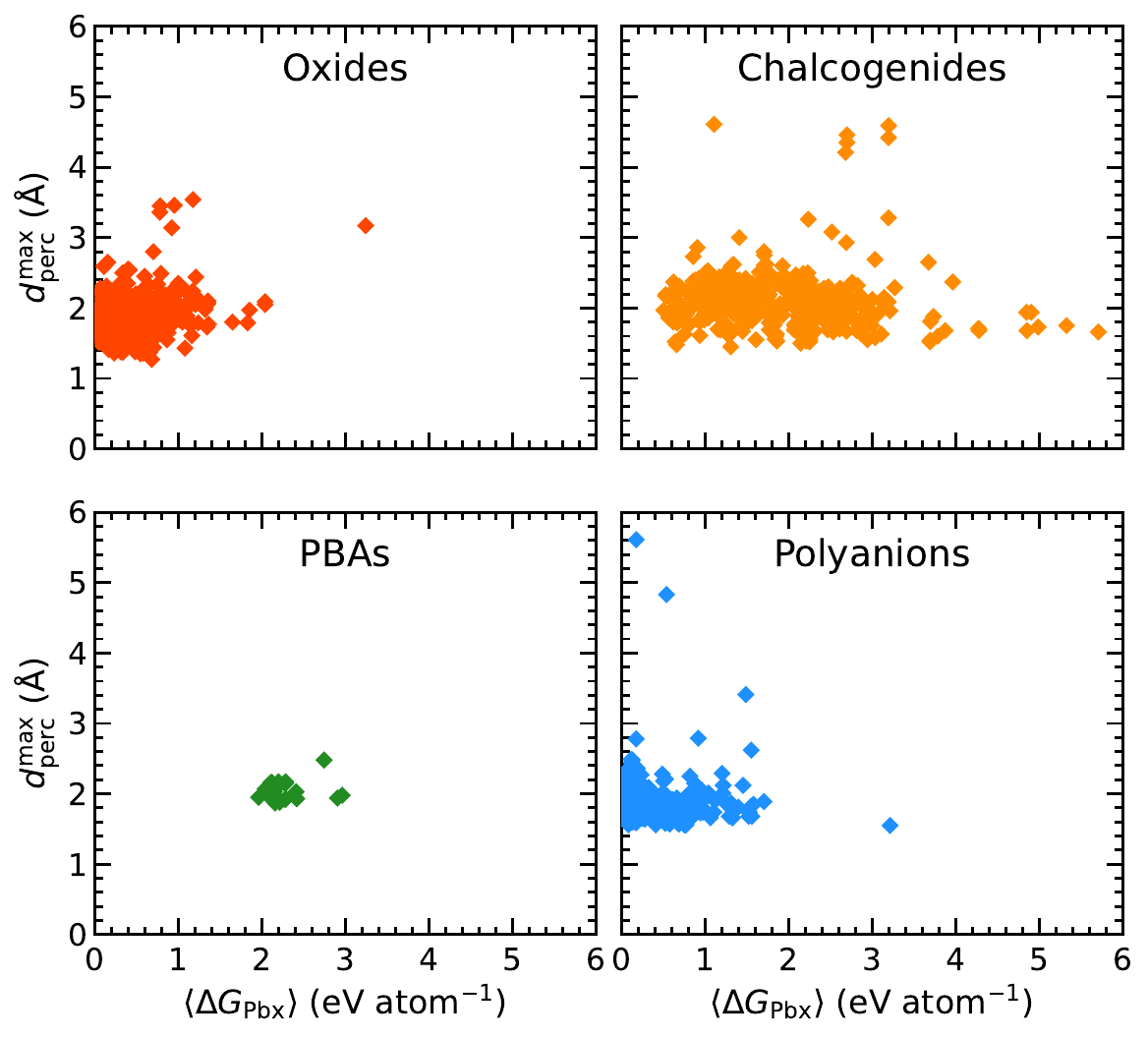}
\caption{\Acf{percdistmax} and \acf{avgdGpbx} results for all screened materials per material class.}
\label{fig:ScrennedMaterialsPerClass}
\end{figure}
\vspace*{\fill}

\vspace*{\fill}
\begin{figure}
\includegraphics[width=0.8\textwidth]{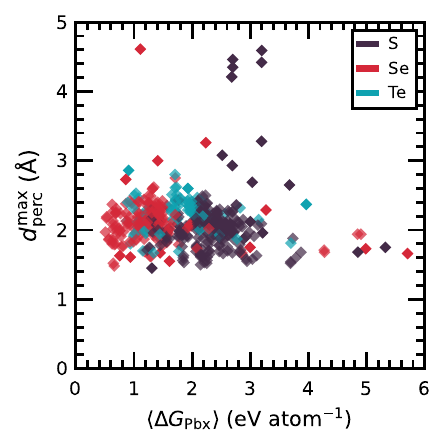}
\caption{\Acf{percdistmax} and \acf{avgdGpbx} for all materials indexed as a chalcogenide with respect to the chalcogenide atom present in the material.}
\label{fig:ScrennedMaterials-ChalcogenideCentre}
\end{figure}
\vspace*{\fill}

\vspace*{\fill}
\begin{figure}
\includegraphics[width=1\textwidth]{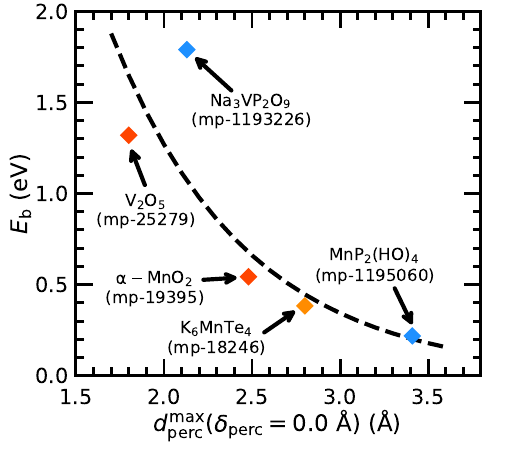}
\caption{Relationship between \acf{percdistmax} and \acf{Eb} for selected materials fitted with an exponential function.}
\label{fig:Void-Eb}
\end{figure}
\vspace*{\fill}

\vspace*{\fill}
\begin{figure}
\includegraphics[width=1\textwidth]{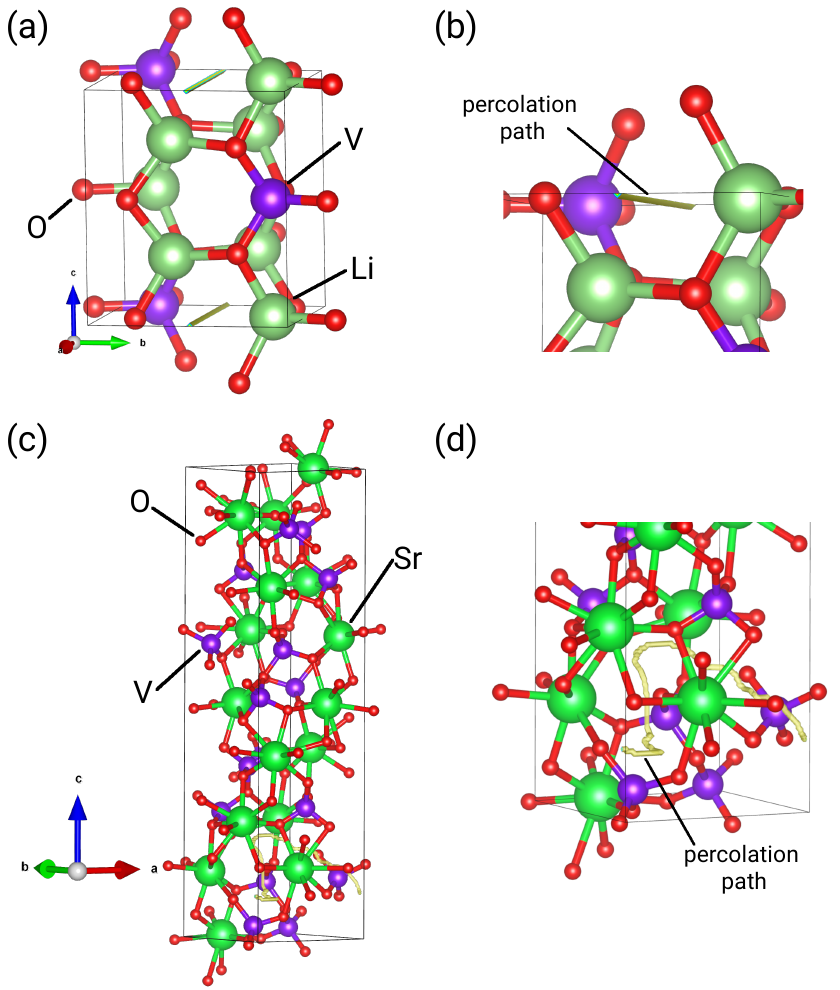}
\caption{Percolation path associated with a \acf{percdistmax} of 1.75~$\text{\AA}$ for (a,b) \ce{Li3VO4} (mp-19219) (\acs{percdisp}~=~0.00~$\text{\AA}$) and (c,d) \ce{Sr2V2O7} (mp-19660) (\acs{percdisp}~=~4.74~$\text{\AA}$).}
\label{fig:PercDisp}
\end{figure}
\vspace*{\fill}

\vspace*{\fill}
\begin{figure}
\includegraphics[width=0.8\textwidth]{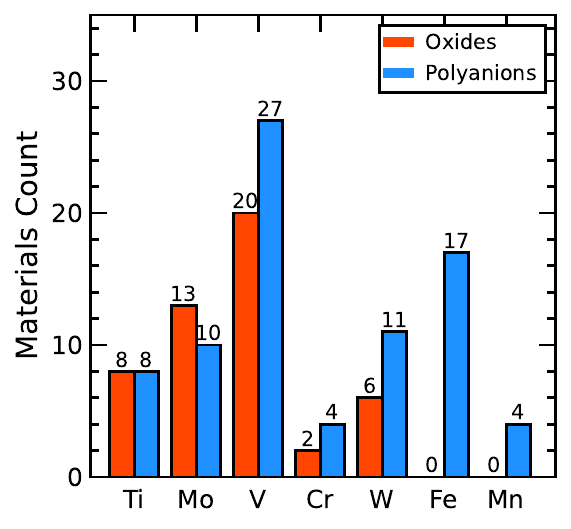}
\caption{Materials selected for \acf{EZn} calculation separated by transition metal centre present in the structure.}
\label{fig:SelectMaterialsPerMetal}
\end{figure}
\vspace*{\fill}

\vspace*{\fill}
\begin{figure}
\includegraphics[width=0.9\textwidth]{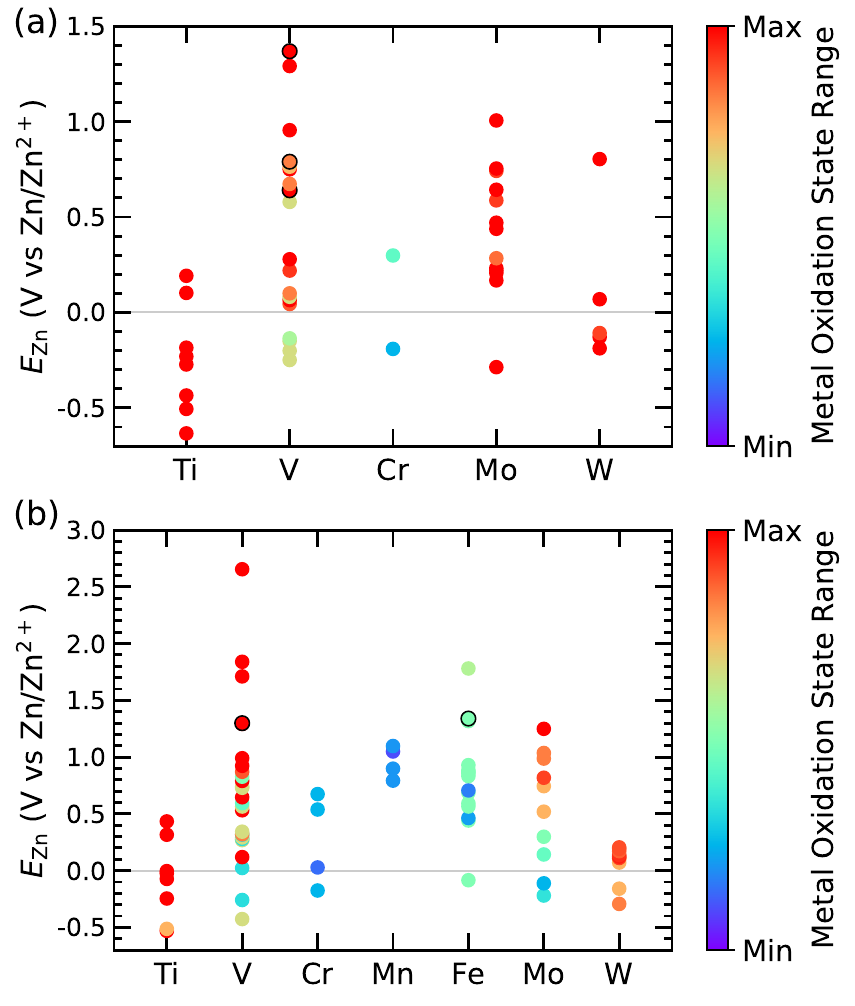}
\caption{\Acf{EZn} results per transition metal centre for (a) oxides and (b) polyanions colour-coded with respect to the lowest and highest experimentally confirmed oxidation state for each transition metal centre.}
\label{fig:PotOxState}
\end{figure}
\vspace*{\fill}

\vspace*{\fill}
\begin{figure}
\includegraphics[width=1\textwidth]{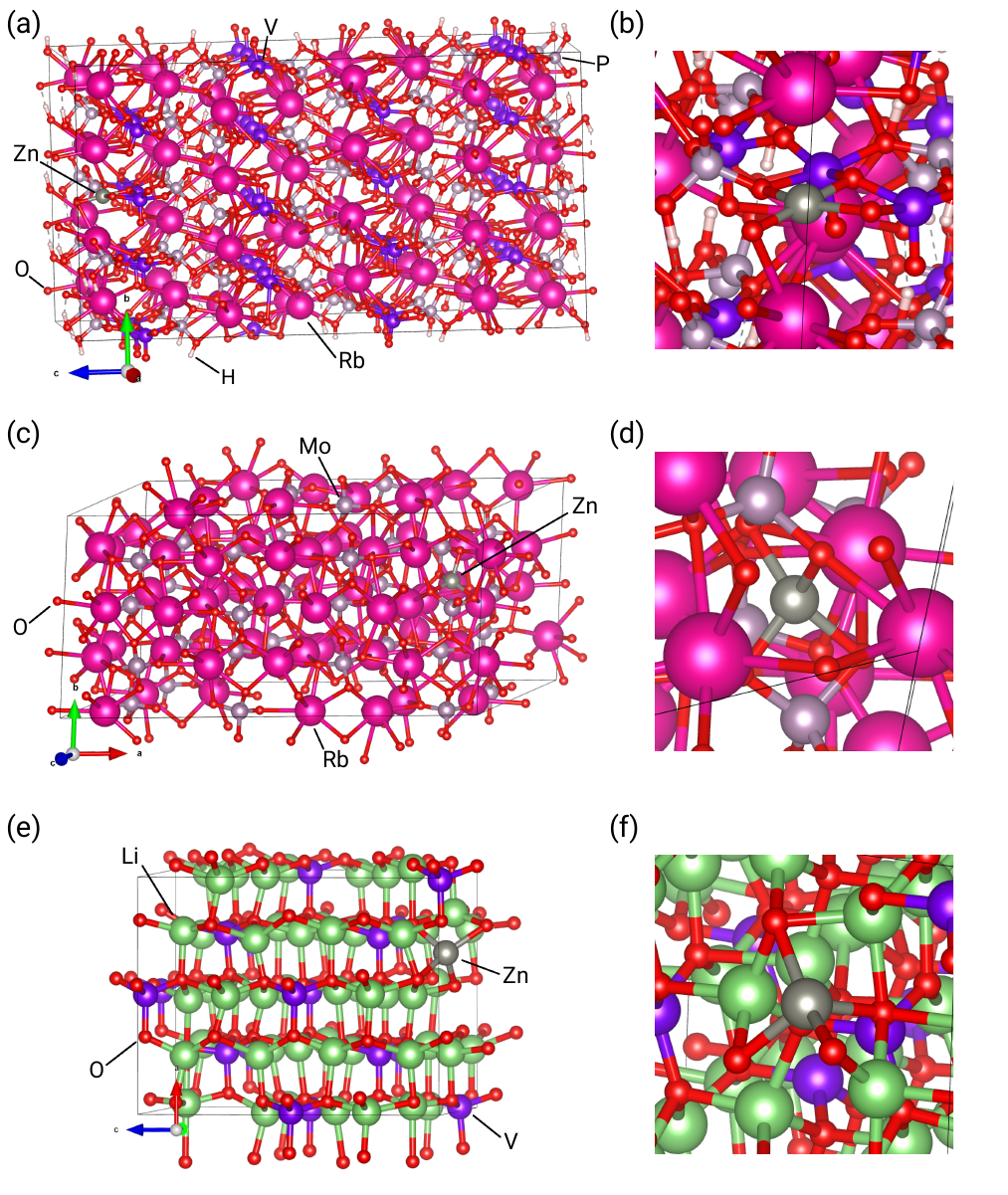}
\caption{Relaxed crystal structure and associated \ce{Zn^{2+}} coordination environment in (a,b) \ce{RbVPHO6} (mp-1201601), (c,d) \ce{Rb2MoO4} (mp-19212), and (e,f) \ce{Li3VO4} (mp-19219).}
\label{fig:CoordStruct-1}
\end{figure}
\vspace*{\fill}

\vspace*{\fill}
\begin{figure}
\includegraphics[width=1\textwidth]{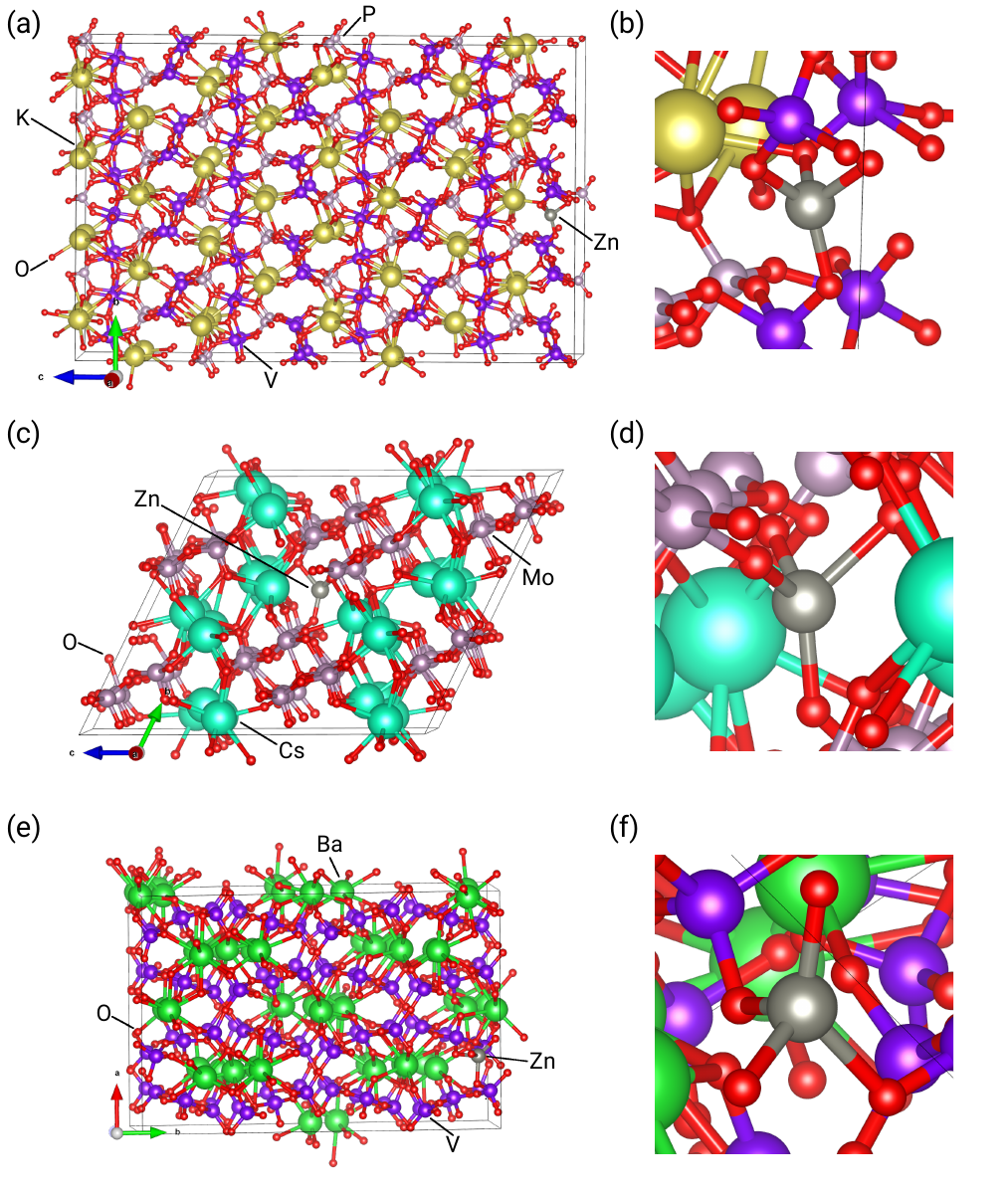}
\caption{Relaxed crystal structure and associated \ce{Zn^{2+}} coordination environment in (a,b) \ce{KV2PO8} (mp-557947), (c,d) \ce{Cs2Mo4O13} (mp-1202511), and (e,f) \ce{BaV2O6} (mp-18929).}
\label{fig:CoordStruct-2}
\end{figure}
\vspace*{\fill}

\clearpage

\vspace*{\fill}
\setlength\extrarowheight{2pt}
\setlength{\tabcolsep}{3pt}
\begin{table}
  \caption{Shortlisted materials calculated \acf{EZninitial}, \acf{EZnavg}, \acf{avgdGpbx}, \acf{Qw}, and \acf{Ww}.}
  \label{tbl:ProposedMaterials}
  \begin{tabular}{c|cccc}
  \hline
  \multirow{2}{*}{Material}  &~~~~\acs{EZninitial}~~~~~\acs{EZnavg}~~~& \acs{avgdGpbx} & \acs{Qw} & \acs{Ww} \\ 
 &  (V~vs~\ce{Zn}/\ce{Zn^{2+}})  & (eV~atom$^{-1}$) & (mA~h~g$^{-1}$) & (W~h~kg$^{-1}$) \\ \hline
\ce{CaFe5P5O22} (mp-818432)	& ~~~~1.78 ~~~~~ 1.30 ~~~ & 0.17 & 162.2 & 210.9 \\
\ce{RbV(PO4)2} (mp-818600)	& ~~~~1.71 ~~~~~ 1.50 ~~~ & 0.19 & 82.1 & 123.2 \\
$\alpha$-\ce{FePO4} (mp-19109) & ~~~~1.34 ~~~~~ 0.96 ~~~ & 0.06 & 177.7 & 170.6 \\
$\beta$-\ce{FePO4} (mp-19752) & ~~~~1.32 ~~~~~ 0.97 ~~~ & 0.06 & 177.7 & 171.6 \\
\ce{KV2PO8} (mp-557947) & ~~~~1.30 ~~~~~ 0.77 ~~~ & 0.11 & 178.7 & 138.3 \\
\ce{SrV2O6} (mp-19038) & ~~~~1.29 ~~~~~ 0.67 ~~~ & 0.13 & 187.8 & 125.0 \\
\ce{Mo2P2O11} (mp-636950) & ~~~~1.25 ~~~~~ 0.96 ~~~ & 0.15 & 124.7 & 119.9 \\
\hline
\end{tabular}
\end{table}
\vspace*{\fill}

\vspace*{\fill}
\begin{figure}
\includegraphics[width=0.8\textwidth]{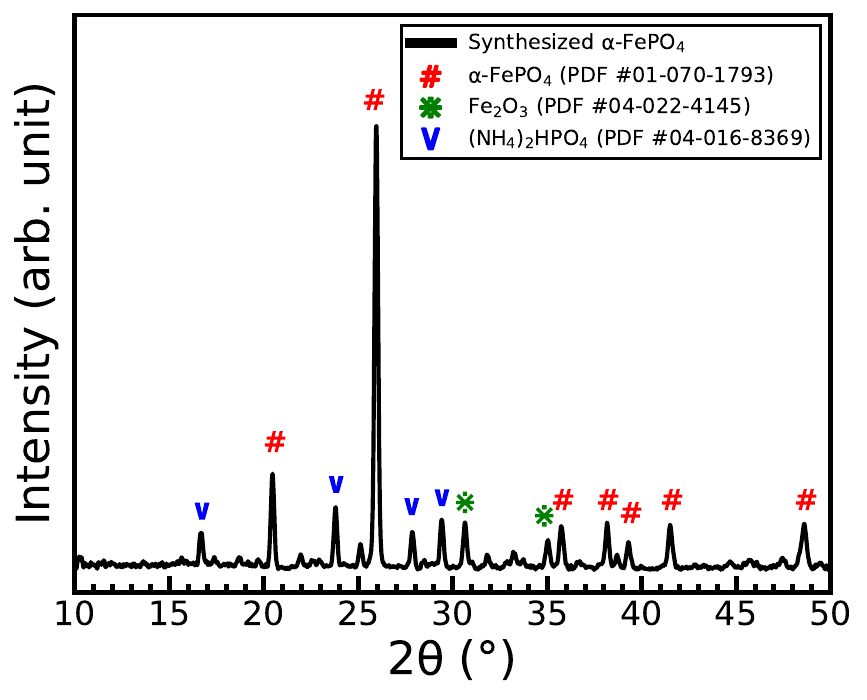}
\caption{\Acf{XRD} diffractrogram for the synthesized $\alpha$-\ce{FePO4} sample. The peaks identified for the target product ($\alpha$-\ce{FePO4}) and precursor impurities (\ce{Fe2O3} and \ce{(NH4)2HPO4}) are also labelled, alongside the numbers for the indexed patterns.}
\label{fig:XRD}
\end{figure}
\vspace*{\fill}

\vspace*{\fill}
\begin{figure}
\includegraphics[width=0.8\textwidth]{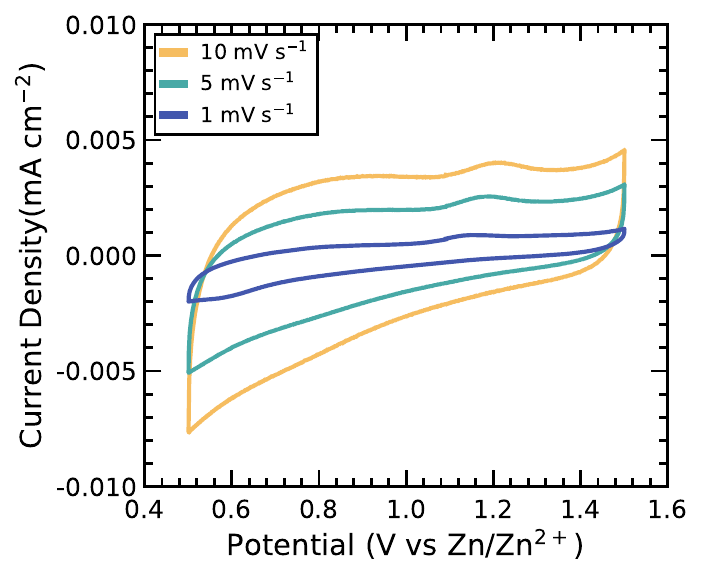}
\caption{\Acf{CV} results for the $\alpha$-\ce{FePO4} electrode at scan rates of 1, 5, and 10~mV~s$^{-1}$, tested in a coin cell with 1~M \ce{ZnSO4} electrolyte and zinc anode.}
\label{fig:CV}
\end{figure}
\vspace*{\fill}

\vspace*{\fill}
\begin{figure}
\includegraphics[width=0.8\textwidth]{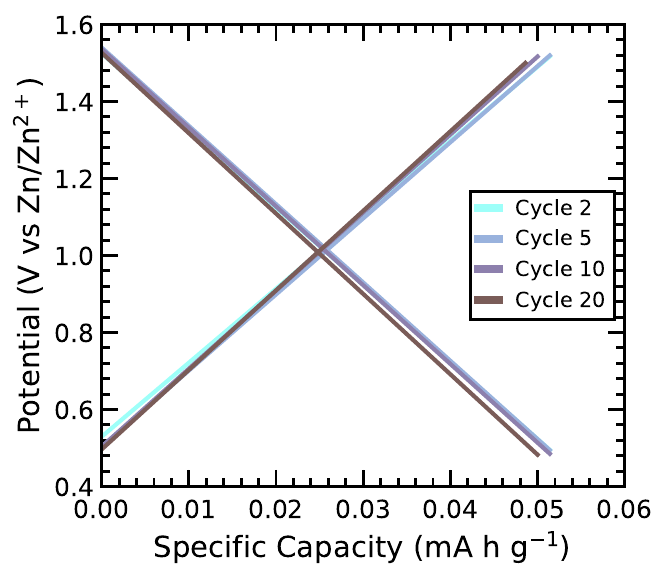}
\caption{\Acf{GCD} results for the $\alpha$-\ce{FePO4} electrode with a current density of 50~mA~g$^{-1}$, tested in a coin cell with 1~M \ce{ZnSO4} electrolyte and zinc anode.}
\label{fig:GCD}
\end{figure}
\vspace*{\fill}

\vspace*{\fill}
\begin{figure}
\includegraphics[width=0.8\textwidth]{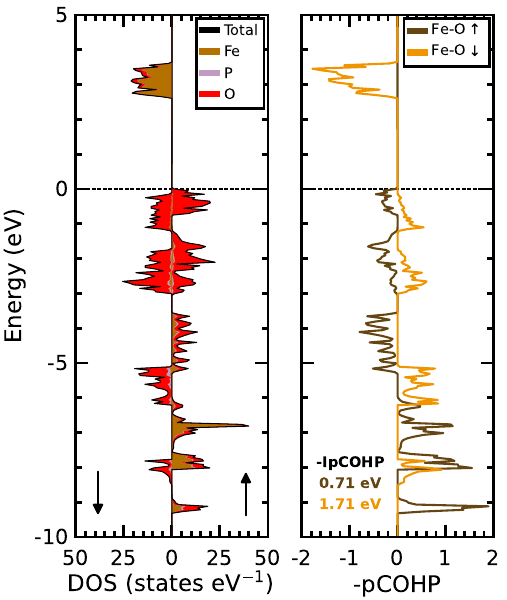}
\caption{\Acf{DOS} and \acf{PDOS} results for $\beta$-\ce{FePO4} (mp-19752), with the associated \acf{-IpCOHP} results for the \ce{Fe}-\ce{O} bond in the structure. Energy is shifted so the \acf{Ef} is at 0~eV. Inset gives the average \acf{-IpCOHP} per \ce{Fe}-\ce{O} bond.}
\label{fig:olivine-FePO4_COHP}
\end{figure}
\vspace*{\fill}

\newpage

\bibliography{bibliography}